\documentclass[%
 reprint,
superscriptaddress,
 amsmath,amssymb,
 aps,
]{revtex4-2}

\usepackage{graphicx}
\usepackage{dcolumn}
\usepackage{bm}

\usepackage{physics}
\usepackage{upgreek}
\usepackage{gensymb}
\usepackage{units}
\usepackage{float}
\usepackage{multirow}
\usepackage[colorlinks=true,linkcolor=blue,urlcolor=blue,citecolor=blue]{hyperref}
\usepackage[markup=underlined]{changes}

\usepackage{appendix}

\begin{document}


\title{A quantum critical line bounds the high field metamagnetic transition surface in UTe$_2$}

\author{Z. Wu}
\thanks{These authors contributed equally to this work.}
\author{T. I. Weinberger}
\thanks{These authors contributed equally to this work.}
\author{A. J. Hickey}
\affiliation{Cavendish Laboratory, University of Cambridge,\\
 JJ Thomson Avenue, Cambridge, CB3 0HE, United Kingdom}

\author{D. V. Chichinadze}
\affiliation{National High Magnetic Field Laboratory, Tallahassee, Florida, 32310, USA}

\author{D. Shaffer}
\affiliation{Department of Physics, University of Wisconsin-Madison, Madison, Wisconsin 53706, USA}

 \author{A. Cabala}
 \affiliation{Charles University, Faculty of Mathematics and Physics,\\ Department of
Condensed Matter Physics, Ke Karlovu 5, Prague 2, 121 16, Czech Republic}

\author{H.~Chen}
\author{M. Long}
\affiliation{Cavendish Laboratory, University of Cambridge,\\
 JJ Thomson Avenue, Cambridge, CB3 0HE, United Kingdom}

 \author{T. J. Brumm}
\affiliation{National High Magnetic Field Laboratory, Tallahassee, Florida, 32310, USA}

\author{W. Xie}
\author{Y. Ling}
\author{Z. Zhu}
\affiliation{Wuhan National High Magnetic Field Center, Wuhan 430074, China}

\author{Y. Skourski}
\affiliation{Hochfeld-Magnetlabor Dresden (HLD-EMFL),\\
Helmholtz-Zentrum Dresden-Rossendorf, Dresden, 01328, Germany}

\author{D. E.~Graf}
\affiliation{National High Magnetic Field Laboratory, Tallahassee, Florida, 32310, USA}

 \author{V.~Sechovský}
 \author{M. Vali{\v{s}}ka}
 \affiliation{Charles University, Faculty of Mathematics and Physics,\\ Department of
Condensed Matter Physics, Ke Karlovu 5, Prague 2, 121 16, Czech Republic}

\author{G. G. Lonzarich} 
\author{F. M. Grosche}
\author{A. G. Eaton}
 \email{alex.eaton@phy.cam.ac.uk}
\affiliation{Cavendish Laboratory, University of Cambridge,\\
 JJ Thomson Avenue, Cambridge, CB3 0HE, United Kingdom}
 
\date{\today}

\begin{abstract}
\noindent
Quantum critical phenomena are widely studied across various materials families, from high temperature superconductors to magnetic insulators. They occur when a thermodynamic phase transition is suppressed to zero temperature as a function of some tuning parameter such as pressure or magnetic field. This generally yields a point of instability -- a so-called quantum critical point -- at which the phase transition is driven exclusively by quantum fluctuations. Here we show that the heavy fermion metamagnet UTe$_2$ possesses a quantum phase transition at extreme magnetic field strengths of over 70 T. Rather than terminating at one singular point, we find that the phase boundary is sensitive to magnetic field components in each of the three Cartesian axes of magnetic field space. This results in the three-dimensional transition surface being bounded by a continuous ring of quantum critical points, the locus of which forms an extended line of quantum criticality -- a novel form of quantum critical phase boundary. Within this quantum critical line sits a magnetic field-induced superconducting state in a toroidal shape, which persists to fields over 70~T. We model our data by a phenomenological free energy expansion, and show how a three-dimensional quantum critical phase boundary -- rather than a more conventional singular point of instability -- anchors the remarkable high magnetic field phase landscape of UTe$_2$.

\end{abstract}

\maketitle
\section{\label{sec:intro}Introduction}

At absolute zero temperature classical phases of matter are completely frozen in place. By contrast, in emergent electronic phases fluctuations from quantum zero-point motion may persist down to temperature $T =$~0~K. For example, when the critical end point (CEP) of a first-order transition is suppressed to zero temperature, a quantum phase transition~\cite{sachdev_QPT} occurs at a quantum critical end point (QCEP)~\cite{BrandoRevModPhys.88.025006}. The enhancement of magnetic or density fluctuations close to a quantum critical point can provide the pairing glue for unconventional superconducting phases to condense~\cite{Monthoux2007}. For example, $d$-wave superconductivity often accompanies quantum phase transitions involving antiferromagnetism~\cite{michon2019thermodynamic,ronning2017electronic}, whereas proximity to ferromagnetic ordering may promote the formation of exotic $p$-wave superconductive phases~\cite{Montu_UGe2,URhGE_Aoki2001,slootenPhysRevLett.103.097003}. In a magnetic field the spin degree of freedom of $p$-wave superconductors can yield complex multi-component superconducting phase diagrams~\cite{leggett_RevModPhys.47.331,TailleferUPt3RevModPhys.74.235}. In some rare cases, such as the ferromagnetic superconductor URhGe, the application of a magnetic field \textbf{H} has been found to induce superconductivity close to a high-$H$ QCEP~\cite{levy2005magnetic,levy2007acute,yelland2011high}.

The paramagnetic metal UTe$_2$ possesses a remarkably rich phase diagram, with up to six distinct superconductive phases having been reported under various conditions of magnetic field tilt angle, temperature and pressure~\cite{Ran2019Science,Aoki2019,Braithwaite2019,Aoki_UTe2review2022,lewin2023review,aoki2020multiple,Ranfieldboostednatphys2019,Thomas2020,wu2024magneticsignaturesmulticomponentsuperconductivity,honda2023pressure}. At $\mu_0H$~=~0~T a superconducting ground state forms at $T_\text{c} =$~2.1~K~\cite{PhysRevMaterials.6.073401MSF_UTe2,tony2024enhanced,aoki2024molten}. Despite the absence of ferromagnetism~\cite{AzariPRL.131.226504}, this superconductive state exhibits multiple characteristics of odd-parity $p$-wave pairing~\cite{Ran2019Science,Aoki_UTe2review2022,Kinjo_SciAdv23} analogous to its ferromagnetic uranium-based cousins UGe$_2$~\cite{Montu_UGe2}, URhGe~\cite{URhGE_Aoki2001} and UCoGe~\cite{slootenPhysRevLett.103.097003}. The rich multidimensional phase space of UTe$_2$ may further contain charge and pair density wave orderings that have been reported by surface-sensitive probes~\cite{CDWaishwarya2023,PDWgu2023} at low-$H$, while at high-$H$ a first-order metamagnetic (MM) transition to a field-polarized paramagnetic (PPM) state occurs for \textbf{H} $\parallel b$, the hard magnetic direction~\cite{Miyake2019}.


The focus of the present study is the highest-$H$ superconducting phase found to date in UTe$_2$, which is located at extremely high field strengths starting from $\mu_0 H \approx$~40~T and extending up to $\mu_0 H \gtrapprox$~70~T. The identification of this superconducting state was first made by contacted and contactless conductivity measurements in pulsed magnetic fields~\cite{Ranfieldboostednatphys2019}. It was subsequently corroborated by bulk thermodynamic~\cite{LANL_bulk_UTe2} and Hall effect~\cite{helm2024} measurements.

Prior studies have reported this high-$H$ superconductivity to be entirely confined inside the PPM state, nestling against the PPM phase boundary within a narrow angular range for \textbf{H} tilted in the crystallographic $b-c$ plane~\cite{Ranfieldboostednatphys2019,knafo2021comparison,lewin2023review,Miyake2021,LANL_bulk_UTe2,helm2024}. Recent measurements have also confirmed its existence for a component of \textbf{H} tilted away from the $b-c$ plane towards the $a$-axis~\cite{lewin2024halo} -- but again, this exotic high-$H$ superconductivity was found exclusively inside the PPM state. The MM transition into the PPM state has been reported to be strongly first-order at low temperatures for the orientations measured so far~\cite{Miyake2019,Miyake2021,Knebel2019,knafo2019magnetic,miyake2022magnetovolume,lewin2023review,helm2024,LANL_bulk_UTe2,valiska2024dramatic,Knebel-caxis-PRB24}. The shape and extent of the MM transition surface, which separates paramagnetic from field-polarized parts of the high field phase diagram, thereby plays an important role in understanding the drivers for the unusual high field superconductivity in UTe$_2$.  Here, we map out the transition surface, show that it is bounded by a critical line, and that the high-$H$ superconducting phase in turn is anchored against that line. These findings suggest a central role for metamagnetic fluctuations in the pairing interaction in UTe$_2$ at high fields.

\section{\label{sec:results}Results}


\subsection{Quantum criticality in intense magnetic fields}
\label{resultsA}

\noindent
We probed the high-$H$ phase diagram of UTe$_2$ with contacted and contactless electrical conductivity measurements on high quality samples grown by a salt flux method~\cite{PhysRevMaterials.6.073401MSF_UTe2,Eaton2024,tony2024enhanced,aoki2024molten}. Figure~\ref{fig:MMdeath} presents contactless magnetoconductance data obtained by the proximity detector oscillator (PDO) technique (see Appendix \ref{appx_exp-tech} for details) in pulsed magnetic fields. For \textbf{H}~$\parallel b$ the MM transition of UTe$_2$ occurs at a characteristic field strength of $\mu_0 H_m \approx 34$~T. As the orientation of \textbf{H} is rotated away from $b$, $H_m$ increases~\cite{Ranfieldboostednatphys2019,helm2024,lewin2024halo}.

This can be seen in Fig.~\ref{fig:MMdeath}c, in which we plot the PDO frequency shift $\Delta f_{\text{PDO}}$ for incremental rotations in the plane that connects the $a$-axis ($\theta_a =$~90$\degree$, $\theta_{b-c} =$~0$\degree$) to the $b-c$ plane at $\theta_{b-c} =$~17$\degree$, $\theta_a =$~0$\degree$, at temperature $T=$~1.7~K. (See Fig.~\ref{fig:wedge-sketch} for a description of this rotation sequence.) Close to $b$, the MM transition manifests as an abrupt increase in $\Delta f_{\text{PDO}}$, due to the dramatic increase in resistivity that occurs upon entering the PPM state~\cite{knafo2019magnetic}. As the orientation of \textbf{H} is tilted further away from the $b-c$ plane towards the $a$ axis, $H_m$ migrates to higher $H$. Notably, the abruptness of the transition gradually diminishes as $H_m$ increases: whereas at $\theta_a =$~6$\degree$ the jump in $\Delta f_{\text{PDO}}$ is very sharp, at $\theta_a =$~17$\degree$ the MM transition is markedly softer, and by $\theta_a =$~20$\degree$ the magnetoconductance curve is very smooth throughout, with no feature indicating the presence of metamagnetism occurring at this angle and temperature for $\mu_0 H \leq$ 78~T.

This softening of the MM transition, as a function of rotation angle at fixed (low) temperature, is very similar to the behavior that occurs as a function of temperature at the fixed orientation of \textbf{H}~$\parallel b$. In Fig.~\ref{fig:MMdeath}e we plot this evolution of $\Delta f_{\text{PDO}}$ as the CEP of the transition is crossed upon raising $T$. The CEP of UTe$_2$'s MM transition for \textbf{H}~$\parallel b$ is reached at $T \approx$~10~K; above this temperature the transition evolves into a broad crossover~\cite{lewin2023review}. For this orientation, at $T = $~1.5~K the jump in $\Delta f_{\text{PDO}}$ is very sharp, exhibiting a large discontinuity in the derivative of the signal with respect to field (panel f). At $T =$~9.0~K there is still a sharp discontinuity in $\nicefrac{\partial f}{\partial H}$; however, at 12~K $\nicefrac{\partial f}{\partial H}$ is much smoother, having crossed the CEP. By contrast, for \textbf{H} tilted 42$\degree$ from [010] towards the [101] direction (labeled as $\theta_{b-ac}$, lower parts of panels e \& f), for all temperatures down to 1.5~K the jump in $\nicefrac{\partial f}{\partial H}$ is notably smoother than that observed at 9.0~K for \textbf{H}~$\parallel b$. This implies that the CEP is located at much lower $T$ for \textbf{H} tilted to $\theta_{b-ac} = 42\degree$ compared to when \textbf{H}~$\parallel b$. In combination, these measurements of the suppression of the MM transition -- for variable tilt angles at fixed temperature, and at incremental temperatures for fixed orientation -- demonstrate that the CEP of the UTe$_2$ normal paramagnetic to polarized paramagnetic phase transition is suppressed towards a QCEP when \textbf{H} is rotated away from the $b$-axis. We sketch a schematic overview of this phenomenon in Fig.~\ref{fig:MMdeath}a,b.

In Figure~\ref{fig:4panel-MM} we track the evolution of UTe$_2$'s MM transition through four additional rotation planes: from $b$ to $a$, $b$ to $ac$, $b$ to $c$ and at a fixed tilt of 7$\degree$ from $b$ to $c$ subsequently rotated towards $a$. For each of these rotation planes we observe the same behavior as in Fig.~\ref{fig:MMdeath}c -- namely, that as \textbf{H} is tilted away from $b$, the sharpness of the MM transition decreases as $H_m$ increases, characteristic of the CEP being suppressed towards a QCEP. This result is notable, when considering the geometrical properties of the PPM state's phase boundary. Figs.~\ref{fig:MMdeath}~\&~\ref{fig:4panel-MM} have shown the sensitivity of the MM transition to rotations of the orientation of \textbf{H}. These experiments were performed by incrementally rotating samples about an axis of rotation; hence we presented the data in terms of the angle of the relative orientation of \textbf{H} with respect to the crystallographic axes. Consider instead the vector \textbf{H} decomposed into its three orthogonal (Cartesian) bases -- i.e. ($H_a,H_b,H_c$). In this three-dimensional magnetic field space, the MM transition surface that separates the normal paramagnetic and polarized paramagnetic phases of UTe$_2$ is defined as the connected surface of points that is swept out by rotating \textbf{H} through each location of $\vb{H_{\textit{m}}}$~\footnote{Note that $\vb{H_{\textit{m}}}$ is itself a function of ($H_a,H_b,H_c$).}. In other words, the MM transition surface is the connected surface of points on which each point lies a distance $\abs{\vb{H_{\textit{m}}}}$ from the origin.

In Fig.~\ref{fig:3D_PD}a we schematically plot the MM transition surface (in orange) on Cartesian magnetic field axes, alongside each of the three (ambient pressure) superconductive phases of UTe$_2$. We label these in ascending order of their upper critical fields as SC1, SC2 and SC3, respectively~\cite{lewin2023review,tony2024enhanced}. For each rotational plane probed by our experiments, we identify the location of that plane's QCEP as the orientation at which the size of the discontinuity in the derivative of the PDO signal tends to zero (marked with green stars), thereby indicating the suppression of the first-order MM phase boundary to zero temperature. The surface marking the first-order transition into the PPM state curves strongly along the $H_a$ direction and is nearly flat along the $H_b$ direction. Our findings show that it is bounded by an undulating quantum critical line, which is indicated in green in Fig.~\ref{fig:3D_PD} -- that we term a quantum critical line (QCL).

\begin{figure}[t!]
    \vspace{0cm}
     \begin{center}
     \includegraphics[width=0.9\linewidth]{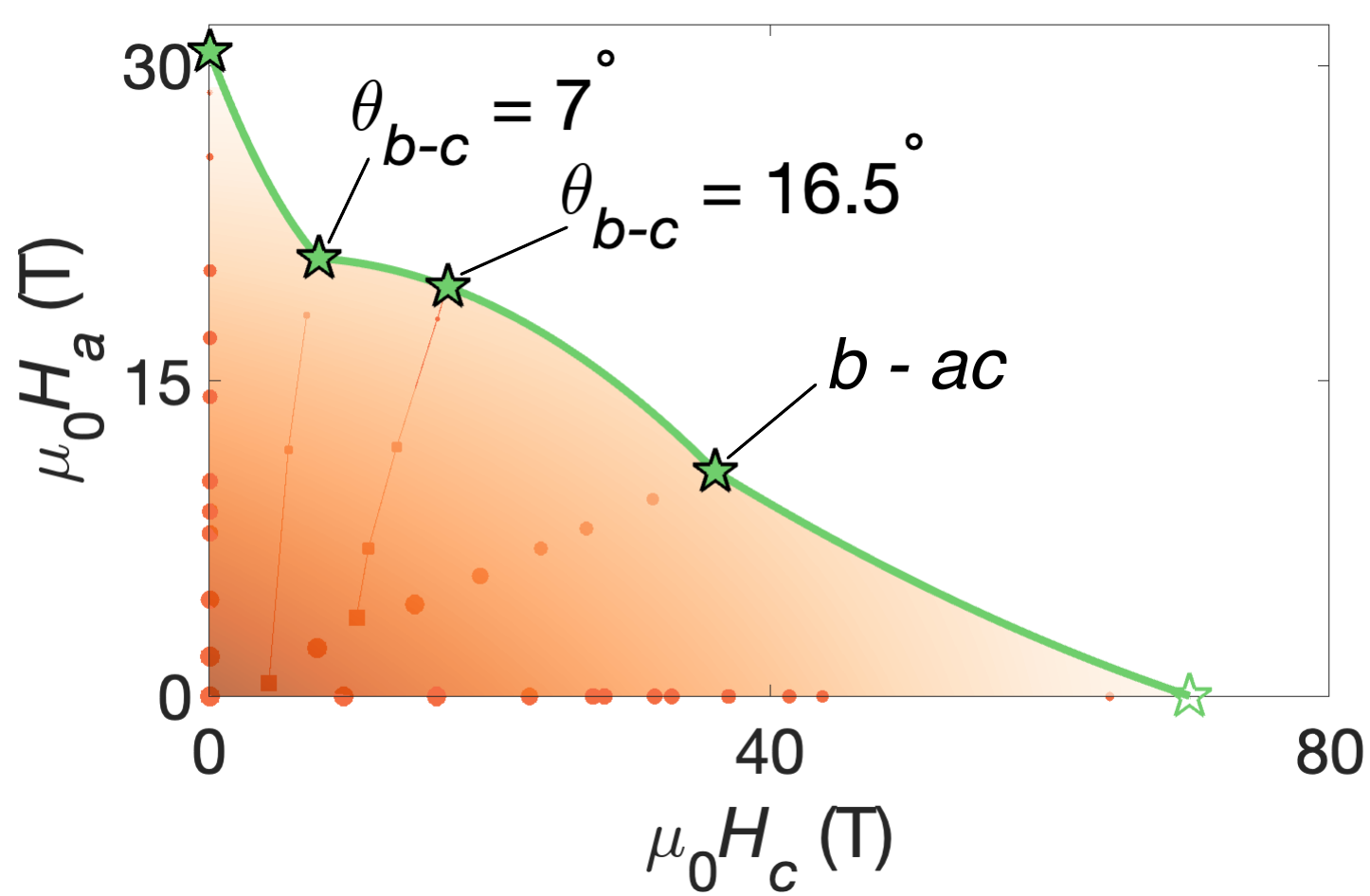} 
     \end{center}
     \vspace{-5pt}
     \caption{Determination of the quantum critical line (QCL). Here we plot the positive quadrant of $H_a - H_c$, viewed looking along the $H_b$ direction, to show how the QCEPs were identified from our rotational magnetoconductance measurements in high magnetic fields. The size of orange points corresponds to the size of the MM transition at that location. The green stars indicate the extrapolation of diminishing MM transition size, to yield the position of the QCEP in each rotation plane. The locus of QCEPs thereby forms the QCL.}
    \label{fig:qclmap}
\end{figure}

In Figure~\ref{fig:qclmap} we show how we constructed this mapping of the three-dimensional QCL of UTe$_2$, by tracking the diminution of the MM transition in each measured rotation plane to obtain the location of that plane's QCEP. We connect these experimentally determined QCEP locations with a smooth green line, to indicate the QCL that separates the polarized and normal paramagnetic states. Inside the region circumscribed by the QCL resides the highest-$H$ superconducting phase of UTe$_2$ (SC3), which wraps around the $b$-axis in a toroidal shape (see Figs.~\ref{pic:SC3-wuhan}~\&~\ref{pic:lewin-comp} and ref.~\cite{lewin2024halo}).

\subsection{Condensation of SC3 before the metamagnetic transition}

In the ferromagnetic superconductor URhGe, $H$-induced superconductivity is found in close proximity to a QCEP at $\mu_0 H \approx$~10~T, identified by the suppression of a first-order MM transition to zero temperature~\cite{levy2007acute}. Our finding of a similar phenomenon in UTe$_2$ -- but in this case stretched along an extended line in three-dimensional \textbf{H}-space at the intense field strengths of 50~T~$< \mu_0 H \lessapprox$~80~T -- motivates a closer examination of SC3 close to the QCL, to which we now turn. 

Figure~\ref{fig:SC3_Spill} shows contacted resistivity measurements by the standard four-terminal technique for a high quality UTe$_2$ specimen (with a residual resistivity ratio of 420, see Fig.~\ref{fig:RRR}). Up to the maximum applied field of the magnet system utilized for this suite of measurements, of 41.5~T, for incremental magnetic field tilt angles rotating from $b$ towards $ac$ at $T =$~0.4~K we suddenly access the SC3 state at $\theta_{b-ac} =$~17.5$\degree$. In Fig.~\ref{fig:SC3_Spill}b we define the purple SC3 region by the observation of zero resistivity $\rho$ (in the low current limit). Interestingly, we observe a broadening of the onset of the SC3 phase for successively higher rotation angles $\theta_{b-ac}$, identified as the turning over of $\rho(H)$ from a positive to a negative gradient in $H$. This is tracked by a dashed black line in Fig.~\ref{fig:SC3_Spill}a. The location of this maximum in $\rho(H)$, which appears to signify the onset of the SC3 state, extends down to a minimum magnetic field of $\approx$~30~T at $\theta_{b-ac} \approx$~34$\degree$, before progressing back up to higher fields for further inclination in the $b-ac$ plane.

\begin{figure}[t!]
    \vspace{0cm}
    \begin{center}
    \includegraphics[width=\linewidth]{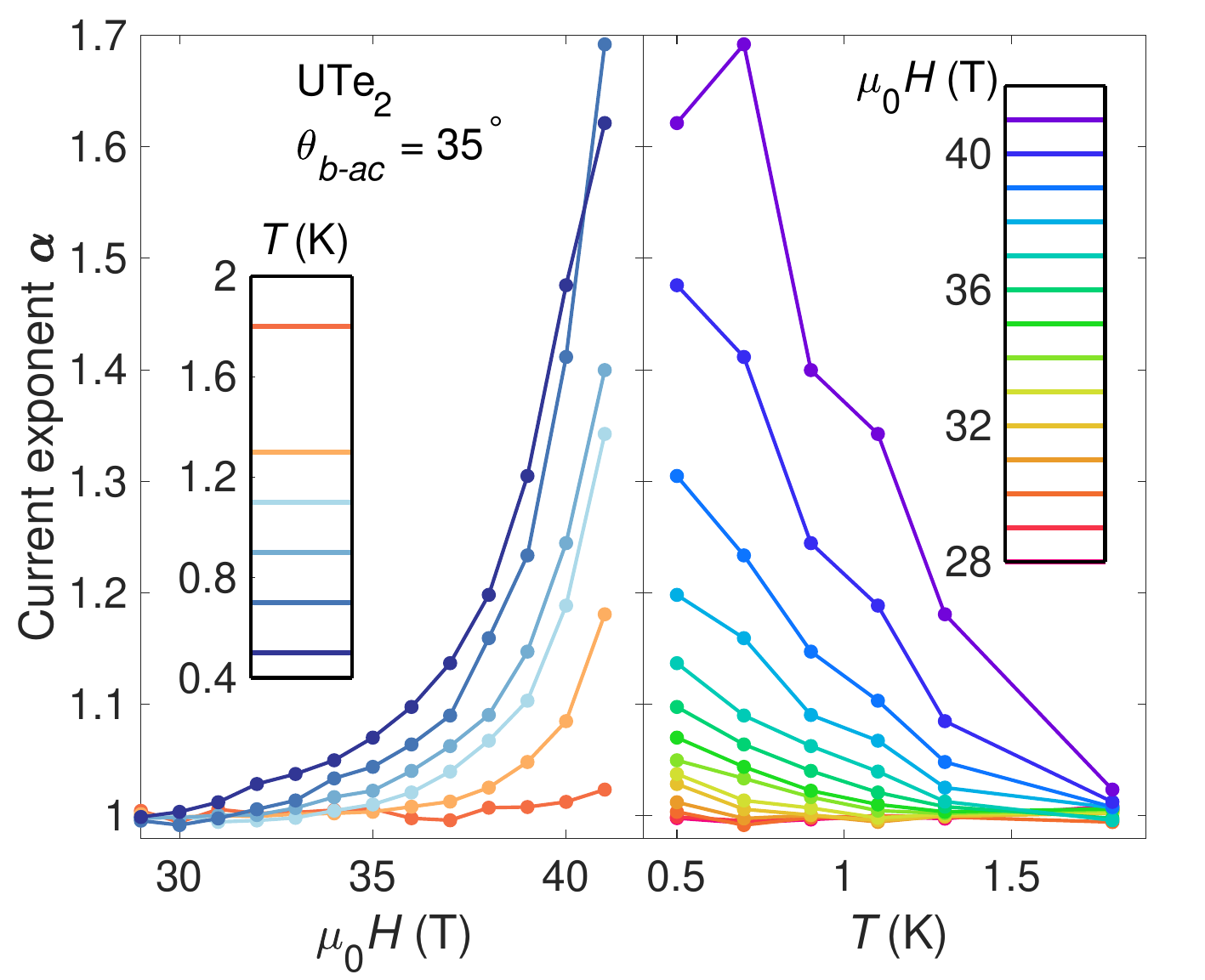} 
    \end{center}
    \vspace{-5pt}
    \caption{Anomalous current--voltage characteristics of SC3. The exponent $\alpha$ extracted from fitting $V \propto J^{\alpha}$ to isothermal field sweeps at incremental currents, plotted as a function of field at incremental temperatures (left) and as a function of temperature for incremental fields (right). The raw data from which this figure is constructed are plotted in Figure~\ref{pic:IV_all}. This analysis indicates that non-Ohmic electrical transport properties -- indicative of the onset of the SC3 state -- onset from $\mu_0H \approx 30$~T for $\theta_{b-ac} =$~35$\degree$, coinciding with the maximum in $\rho(H)$ (in the low $J$ limit) for this orientation.}
    \label{pic:IV_logVlogI}
\end{figure}

This wide magnetic field domain of negatively sloped magnetoresistance exhibits a highly non-linear relationship between excitation current density $J$ and voltage $V$. Whereas standard Ohmic behavior is observed for all measured temperatures at 25~T when the sample is in the normal state, at 41~T a strong deviation is observed at low temperatures (Fig.~\ref{fig:SC3_Spill}e,f). This appears to indicate the occurrence of vortex motion~\cite{fluxflowPhysRev65,BlattRevModPhys94,yu-te2021unconventional}, perhaps due to the presence of low pinning forces in this high quality sample. We note that exotic vortex properties have recently been reported for the SC1 phase at low $H$~\cite{wang2024vortexstripesute2}; whether the SC3 state also possesses anomalous vortex behavior is an open question. In Figure~\ref{pic:IV_logVlogI} we plot the exponent $\alpha$ fitted to $V \propto J^{\alpha}$ to illustrate the departure from Ohmic behavior at low temperatures for $\mu_0H \gtrapprox$~30~T. This non-linearity between $J$ and $V$ clearly extends down far below $H_m$. We therefore identify a broad range of the UTe$_2$ high field phase diagram, for \textbf{H} tilted close to the QCL, to be occupied by the onset of the SC3 phase.

For \textbf{H}~$\parallel b$, the first-order MM transition to the PPM is observed at $\mu_0 H_m \approx$~34.5~T in the zero temperature limit~\cite{Miyake2019,Miyake2021,Aoki2021,tony2024enhanced}. As discussed in Section~\ref{resultsA}, on warming the CEP is crossed at $T \approx$~10~K and the transition then evolves into a broad crossover, characterized by e.g. a maximum in $\rho(H)$, as the PPM gradually melts and $H_m$ decreases until $T \approx$~30~K where $\mu_0 H_m$ reaches 0~T~\cite{Aoki_UTe2review2022,lewin2023review,Miyake2019,Miyake2021,valiska2024dramatic}. Accordingly, we identify the location of $H_m$ for each temperature as the field strength at which $\nicefrac{\partial \rho}{\partial H} = 0$, as shown in Figure~\ref{pic:derivs}.  We schematically plot the corresponding phase landscapes for \textbf{H}~$\parallel b$ (i.e. $\theta_{b-ac} = 0\degree$) and $\theta_{b-ac} = 35\degree$ in Fig.~\ref{fig:SC3_Spill}c. Strikingly, upon warming above $T_c^{SC3}$ at $\theta_{b-ac} = 35\degree$, we do not observe a signature of the PPM state in our accessible field range of $\mu_0 H \leq$~41.5~T until $T \approx 15$~K (Fig.~\ref{pic:derivs}).

Furthermore, we performed a series of rotations in the $b-ac$ plane at 6~K -- to suppress all superconducting states and therefore solely probe the PPM phase -- to map out the boundary of the MM transition in this plane, which we plot as red circles in Fig.~\ref{fig:SC3_Spill}b (see Fig.~\ref{fig:blue-spill} for raw data). The MM transition clearly bisects a substantive region of the SC3 domain (as determined by the observation of zero $\rho$ in the limit of low $J$ and $T$). Therefore, our measurements have uncovered that a portion of the SC3 phase -- and its broad onset region -- extends considerably outside the PPM state in the $b-ac$ plane as \textbf{H} is tilted close to the QCL (Fig.~\ref{pic:3D_data}). This is in stark contrast to prior measurements of SC3 in the $b-c$ plane, comparatively far from the QCL, that observed SC3 exclusively within the PPM host-state~\cite{Ranfieldboostednatphys2019,knafo2021comparison,lewin2023review,Miyake2021,LANL_bulk_UTe2,helm2024,frank2024orphan}.

\subsection{Landau theory of metamagnetic transitions in three dimensions}

The SC3 phase of UTe$_2$, which persists to $\mu_0 H \gtrapprox$~70~T despite its relatively modest $T_c$ of 2.4~K~\cite{tony2025brief}, has posed a formidable theoretical challenge to our understanding of high-$H$ superconductivity~\cite{Aoki_UTe2review2022,lewin2023review}.
As prior high-$H$ measurements in the $b-c$ plane only observed SC3 coexisting alongside the PPM state, the field-induced compensation of magnetic exchange~\cite{JaccPet1962PRL} was proposed as a possible microscopic origin~\cite{helm2024}. However, our finding of the zero resistance SC3 state extending outside the PPM (Fig.~\ref{fig:SC3_Spill}) -- accompanied by an extended field region of the onset of SC3 even further away from the MM phase transition line upon approaching the QCL -- argues strongly against this scenario.

To shed light on the origin of the QCL, we model the MM transition using a Landau model, including terms up to sixth order in the magnetic order parameter \(\mathbf{M}\), the magnetization:
\begin{align}
    \mathcal{F}[\mathbf{M}](\mathbf{H})&=\frac{1}{2}\chi^{-1}_{i}M_i^2+\frac{1}{4}\beta_{ij} M_i^2M_j^2\\ &+ \frac{1}{6} \delta_{ijk} M_i^2 M_j^2 M_k^2 -\mathbf{M\cdot H}
\label{F_metamagnetic}
\end{align}
where \(i,j,k=a,b,c\) and \(\chi^{-1}_{i}, \beta_{ij}\), and \(\delta_{ijk}\) are model parameters (see Appendix~\ref{app-theory})~\cite{Yamada_metamagnetic,Mineev15,LinNevidomskyyPaglione20}.
The MM phase transition occurs when the free energy has two local minima at $\mathbf{M}=\textbf{M}_{0}$ and \(\mathbf{M}=\mathbf{M}_*\) corresponding to the normal paramagnetic and PPM phases, respectively. With \(\beta_{bb}<0\), the \(H_b\) component of the external field tends to lower the minimum at \(\mathbf{M}_*\) faster than the minimum at \(\mathbf{M}_0\), resulting in the MM transition when \(\mathcal{F}(\mathbf{M}_*)=\mathcal{F}(\mathbf{M}_0)\), which generically defines a 2D surface in the 3D space of \(\mathbf{H}\).

In general, the model parameters are temperature dependent, such that as temperature increases the two minima merge at a CEP. For simplicity, here we focus on the low temperature limit and consider the effects of \(H_a\) and \(H_c\). Due to the mixed terms \(\beta_{ij}\) and \(\delta_{ijk}\), both \(H_a\) and \(H_c\) (which induce \(M_a\) and \(M_c\)) in effect indirectly modify the \(\chi_{b}\) and \(\beta_{bb}\) coefficients and so act in a similar way to the temperature. As a result, large \(H_a\) and \(H_c\) tend to move \(\mathbf{M}_0\) and \(\mathbf{M}_*\) towards each other, until they merge at the critical end line (at finite temperature) or the QCL (at zero temperature) bounding the MM phase transition surface.

We confirm this general picture by numerically minimizing the free energy of Eqn.~\eqref{F_metamagnetic}, determining the MM transition surface and the QCL as a function of \textbf{H}, as shown in Figure~\ref{fig:QCLsim}. With the given parameters (see Appendix~\ref{app-theory}) the model matches the experimental data from Section~\ref{resultsA} reasonably well and captures the main qualitative features. Assuming SC3 is driven by the fluctuations in the vicinity of the QCL, the shape of the QCL is qualitatively close to the observed field direction dependence of SC3, with orbital pair-breaking and spin effects potentially also relevant.
The QCL fluctuation scenario is further supported by the fact that, analogous to the case of ferromagnetic uranium-based superconductors such as URhGe~\cite{levy2007acute,yelland2011high,Mineev15}, SC3 should be expected to occur on both sides of the QCL as we observe experimentally (Fig.~\ref{fig:SC3_Spill}). 



\section{Discussion}
In combination, our high-$H$ measurements of UTe$_2$ presented in Section~\ref{sec:results} (i) identify that an extended line of QCEPs bounds the surface of first-order MM transitions in three-dimensional \textbf{H}-space at the remarkably high magnetic field scale of over 70~T, (ii) anchor high-$H$-induced superconductivity in UTe$_2$ close to this boundary, and (iii) reveal that in a narrow angular range close to the QCL the onset of the SC3 state extends down to fields that lie far below the MM transition field.

We note that the concept of a QCL has previously been considered in holographic theoretical studies of quantum criticality~\cite{gouteraux2013quantum,kim2012holographic}. Additionally, extended lines of quantum criticality have been reported in experimental investigations of the pressure and doping phase diagrams of several cerium-based heavy fermion superconductors~\cite{park2006hidden,RonningPRB06,ZaumPRL10-QCL,chang2023scaled}. For example, in the case of CeCoIn$_5$, the QCL of this material is a function of pressure $p$ and $H$~\cite{ZaumPRL10-QCL}. The QCL connects a series of quantum critical points located at the edge of antiferromagnetic order, which trace out a smooth curve when plotted in two-dimensional $p-H$ space. By contrast, in the present study we find that the sensitivity of the termination of UTe$_2$'s PPM state to each of the three orthogonal directions of \textbf{H} leads to a three-dimensional QCL at ambient pressure. Given that the field scale of both SC3 and the MM phase boundary have been reported to decrease with increasing $p$~\cite{LinNevidomskyyPaglione20,ran2021expansion}, it is possible that under compression the QCL of UTe$_2$ would be extruded to form a quantum critical surface in the four-dimensional space of $p$-\textbf{H}. Further high-$H$ measurements under pressure are required to test this hypothesis.

Our findings here -- of $H$-induced superconductivity in UTe$_2$ being located in proximity to QCEPs of a first-order MM phase boundary -- mirror prior studies on the related compound URhGe~\cite{levy2005magnetic,levy2007acute,levy2009JPCM,yelland2011high,Aoki_ferro_review2019}. This material possesses a ferromagnetic groundstate, with moments oriented along the $c$-axis. For \textbf{H} applied along the $b$ direction, a metamagnetic transition occurs at $\mu_0H_b \approx$~12~T, which can be interpreted as a rotational transition of the orientation of magnetic moments from $\parallel c$ to $\parallel b$~\cite{levy2005magnetic}, with a corresponding change of the easy magnetic axis~\cite{Aoki_ferro_review2019}. $H$-induced superconductivity is observed at low temperatures over the interval of 8~T~$\lessapprox \mu_0H_b \lessapprox$~13~T~\cite{levy2005magnetic}. This magnetic field domain lies predominantly within the ferromagnetic phase, with a small portion of the superconductivity spilling out to the other side of the MM transition. Applying an $a$-axis magnetic field component has very little effect on the $H$-induced superconductivity, which can survive up to (at least) $\mu_0\abs{\textbf{H}}=$~32~T~\cite{levy2009JPCM} (where here $\abs{\textbf{H}} \equiv \sqrt{H_a^2 + H_b^2}$). However, at $\mu_0H_b =$~12~T, just a small $c$-axis field component of $\mu_0 H_c <$~2~T quickly destroys the $H$-induced superconductivity, as the QCEP of the first-order MM phase boundary is crossed~\cite{levy2007acute,levy2009JPCM,Aoki_ferro_review2019}. Nuclear magnetic resonance (NMR) measurements have provided microscopic evidence that the $H$-induced superconductivity is mediated by quantum critical magnetic fluctuations stemming from the suppression of the first-order MM phase boundary to zero temperature~\cite{TokunagaPRL15,TokunagaPRB16}.

\begin{figure*}[h!]
\vspace{-2.5cm}
\begin{center}
\includegraphics[width=1\linewidth]{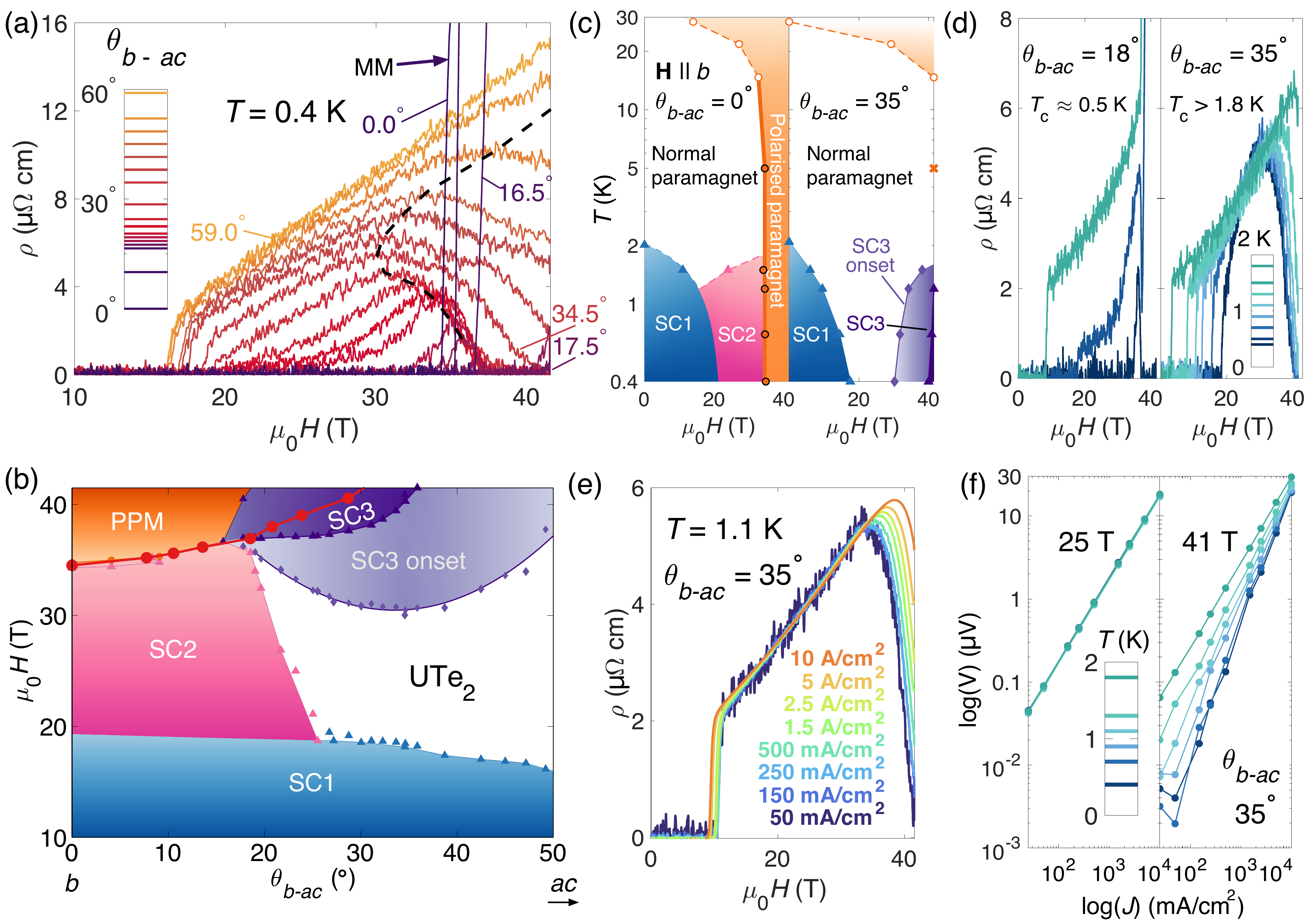}
\end{center}
\caption{Spillover of SC3 outside the polarized paramagnetic phase. (a) Electrical resistivity $\rho(H)$ at successive magnetic field tilt angles in the $b-ac$ rotation plane at $T =$~0.4~K. At $\theta_{b-ac} =$~17.5$\degree$ zero resistance is observed over the field interval of 0~T~$\leq \mu_0 H \leq$~41~T as each of SC1, SC2 and SC3 are successively accessed. For ease of presentation only selected curves are plotted -- see Fig.~\ref{fig:blue-spill} for the full angular dataset. (b) The phase landscape of UTe$_2$ plotted radially in $\theta_{b-ac}$. An extended region of the onset of SC3 is observed outside the PPM state, identified by the turning over from positively to negatively sloped $\rho(H)$ marked by the black dashed line in panel (a). The red circles defining the PPM boundary were measured above $T_c^{SC3}$ (see Fig.~\ref{fig:blue-spill}). (c) The evolution of the SC1, SC2, SC3 and PPM phases for $\theta_{b-ac} =$~0$\degree$ (left) and $\theta_{b-ac} =$~35$\degree$ (right). At $\theta_{b-ac} =$~35$\degree$ the MM transition is not observed at 5~K (marked by a cross -- see Fig.~\ref{pic:derivs}). This shows that SC3 extends outside the PPM state for \textbf{H} tilted close to the QCL. (d) A comparison of the temperature sensitivity of SC3 at $\theta_{b-ac} =$~18$\degree$ (left) and $\theta_{b-ac} =$~35$\degree$ (right). On the edge of the SC3 dome at $\theta_{b-ac} =$~18$\degree$ the $T_{\text{c}}$ is only $\approx$~0.5~K, whereas at $\theta_{b-ac} =$~35$\degree$ SC3 is much more robust to elevated temperatures. (e) Effective resistivity at $\theta_{b-ac} =$~35$\degree$ for incremental excitation current densities as indicated. Data at other temperatures are presented in Fig.~\ref{pic:IV_all}. (f) Log-log plot of voltage $V$ versus current density $J$  from 0.4-1.8~K at $\theta_{b-ac} =$~35$\degree$. Ohmic $J-V$ characteristics are observed in the normal state at $\mu_0 H =$~25~T for all currents, whereas at $\mu_0 H =$~41~T the $J-V$ behavior is highly non-Ohmic. We therefore identify the turning over of the $\rho(H)$ traces in (a) as marking the onset of SC3, characterized by this strongly non-linear relationship between $J$ and $V$.
}
\label{fig:SC3_Spill}
\end{figure*}

\begin{figure}[t!]
    \vspace{-0cm}
     \begin{center}
     \includegraphics[width=1\linewidth]{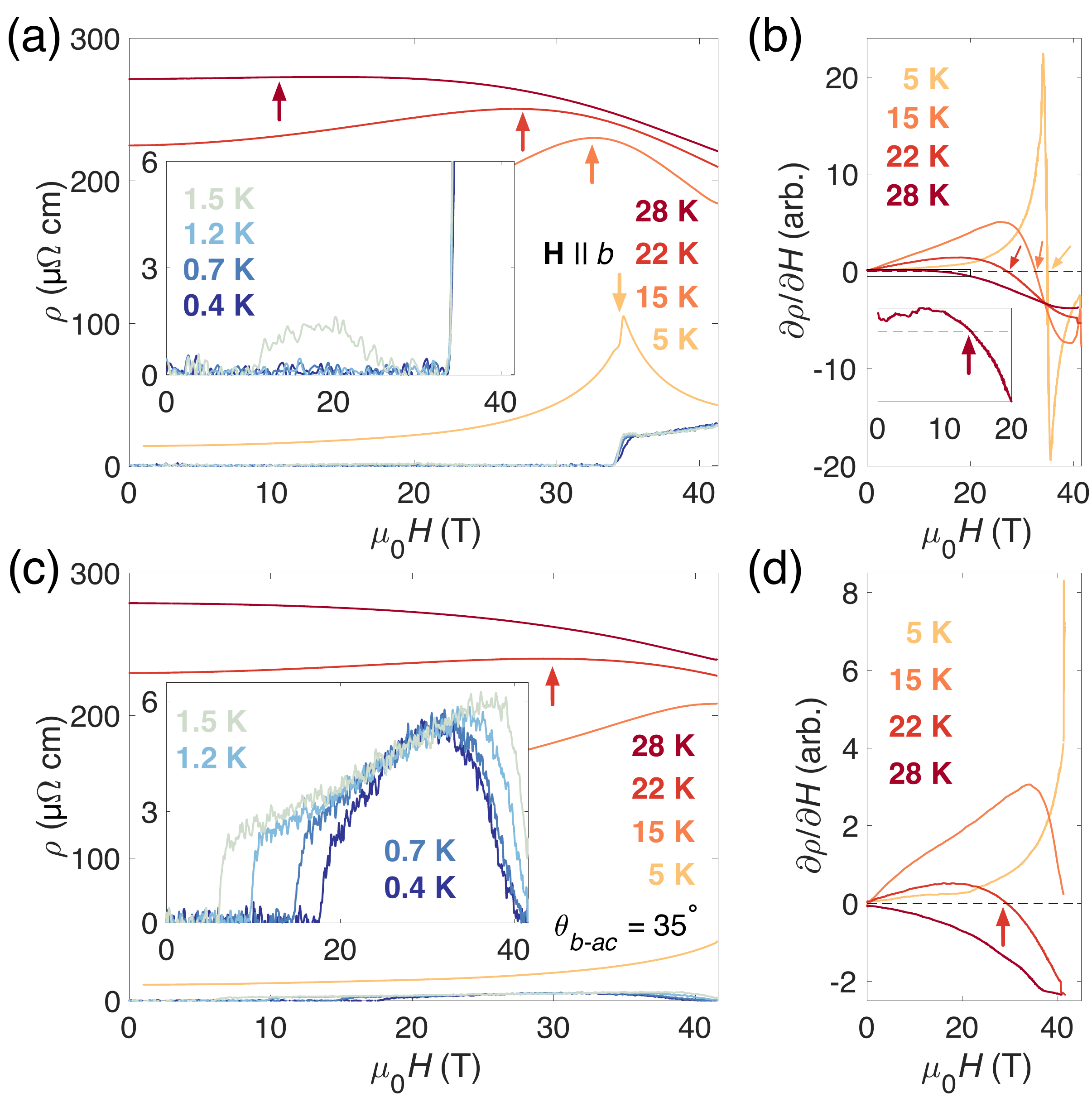} 
     \end{center}
     \vspace{-5pt}
     \caption{(a) $\rho(H)$ for \textbf{H}~$\parallel b$ at incremental temperatures, with corresponding $\nicefrac{\partial \rho}{\partial H}$ in (b). The inset shows the derivative of the 28~K curve. (c) $\rho(H)$ at $\theta_{b-ac}$~=~35$\degree$, with corresponding $\nicefrac{\partial \rho}{\partial H}$ in (d). These data were used to construct Fig.~\ref{fig:SC3_Spill}c. In all panels, arrows mark zero $\nicefrac{\partial \rho}{\partial H}$, identifying the location of the metamagnetic transition. Importantly, zero resistance is observed at $\mu_0 H \approx$~40~T at low temperatures for $\theta_{b-ac} =$~35$\degree$ even though $\nicefrac{\partial \rho}{\partial H}$ rises monotonically at $T$~=~5~K, indicating that in the low temperature limit for this orientation the metamagnetic transition field is located significantly above the maximum accessed field strength of 41.5~T for these measurements -- even though zero resistance is observed at low temperatures. This shows that, in proximity to the QCL, the SC3 phase has a region of spillover outside the PPM state.}
    \label{pic:derivs}
\end{figure}

There are numerous similarities between URhGe and our findings here on UTe$_2$. The MM transition in UTe$_2$ also marks a change in the easy axis -- in this case, from $a$ to $b$~\cite{Miyake2021}. Both materials possess $H$-induced superconductivity in proximity to the QCEPs of a first-order MM phase boundary, the CEP of which may be tuned to absolute zero by the tilt angle of the magnetic field. While the $H$-induced superconductivity of URhGe lies predominantly within the ferromagnetic state, extending a little to higher $H$ on the other side of the MM transition, in UTe$_2$ SC3 resides predominantly within the PPM state, extending a little to lower $H$ on the other side of the MM transition (Fig.~\ref{fig:SC3_Spill}). The key differences between the two materials are the magnetic ground state properties, the associated magnetic and superconducting energy scales, and the dimensionality of their quantum critical phase boundaries.

At first glance, perhaps the most striking contrast between these two materials is the remarkably high magnetic field scales of the metamagnetic and associated quantum critical phenomena underpinning high-$H$-induced superconductivity in UTe$_2$. A commonality shared by both compounds is that their $H$-induced superconducting states possess higher maximal $T_c$ values than is observed for the superconductive phases at ambient magnetic field~\cite{levy2005magnetic}. However, at 2.4~K~\cite{tony2025brief}, the maximal $T_c$ of UTe$_2$ is markedly higher than the comparable value of 0.4~K in the case of URhGe. The location of $\mu_0 H_m$ at 34~T is also considerably higher than URhGe's value of 12~T. Regarding the magnetic properties, at low $H$ UTe$_2$ lacks long-range magnetic order; however, it exhibits clear signatures of antiferromagnetic (high wavevector $Q$) correlations~\cite{Knafo104.L100409,raymond2021feedback,Duan2021AFMfluc}. It is not clear if these fluctuations persist up to the high field strengths at which the QCL and SC3 are found. Given that the quantum critical phase boundary of UTe$_2$ is sensitive to all three orthogonal spatial magnetic field components -- unlike URhGe, in which only two dimensions are relevant -- it is not inconceivable that the SC3 state might be mediated by a combination of low and high $Q$ fluctuations. For \textbf{H}~$\parallel b$, NMR measurements of UTe$_2$ have observed strong ferromagnetic-like longitudinal spin fluctuations for $\mu_0 H_b \gtrapprox$~15~T, which were found to diverge as $H \rightarrow H_m$~\cite{tokunaga2023longitudinal}. These findings appear to concur with an analysis of the magnetic entropy landscape upon approaching the MM transition, in which strong $Q = 0$ fluctuations were discerned~\cite{TokiwaPRB2024}. 

\begin{figure*}[h!]
\vspace{-0cm}
\begin{center}
\includegraphics[width=.8\linewidth]{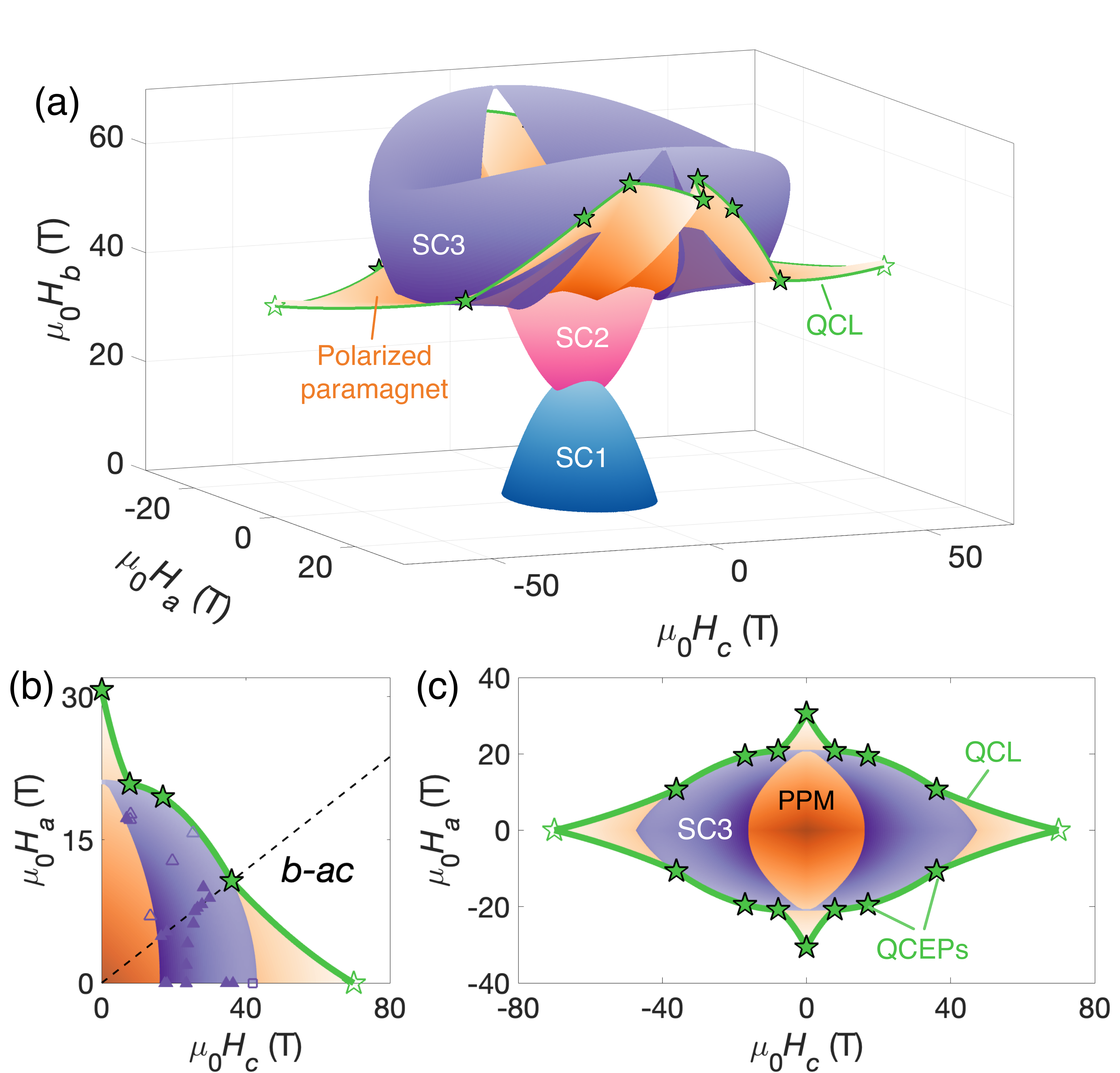}
\end{center}
\caption{The high magnetic field quantum critical phase landscape of UTe$_2$. (a) Superconducting phase diagram of UTe$_2$. We label the superconducting ground state at $\mu_0 H =$~0~T as SC1 (in blue), while the pink region identifies a distinct superconducting state~\cite{Rosuel23} attained for $\mu_0 H_b \lessapprox$~35~T (SC2). The highest-$H$ superconductivity (SC3) is in purple, while the first-order metamagnetic transition surface into the polarized paramagnetic (PPM) state is colored orange. Tilting \textbf{H} away from the $b$-axis induces a quantum phase transition at the edge of the PPM state, identified by quantum critical end points (QCEPs) marked with green stars. Solid green stars are directly measured in this study, while the open green stars are at higher fields and are extrapolated from our 80~T measurements (see Fig.~\ref{fig:bc-extrpol}). The locus of QCEPs thus forms an extended quantum critical line (QCL), drawn in green. (b) The positive quadrant of $H_a$--$H_c$ with data points from pulsed field measurements at $T \approx$~0.7~K. The $b-ac$ rotation, [010]--[101], is marked in black. Solid triangular points are from this study; open triangles are from ref.~\cite{lewin2024halo} with the open square point at $\mu_0 H_a$~=~0~T from ref.~\cite{helm2024}. We use the location of these experimentally determined points to set the boundaries of purple shading throughout this figure. (c) Bird's-eye view along $H_b$ showing how the toroidal shape of the SC3 phase is enveloped by the QCL.}
\label{fig:3D_PD}
\end{figure*}

\begin{figure}[t!]
    \vspace{0cm}
    \begin{center}
    \includegraphics[width=1\linewidth]{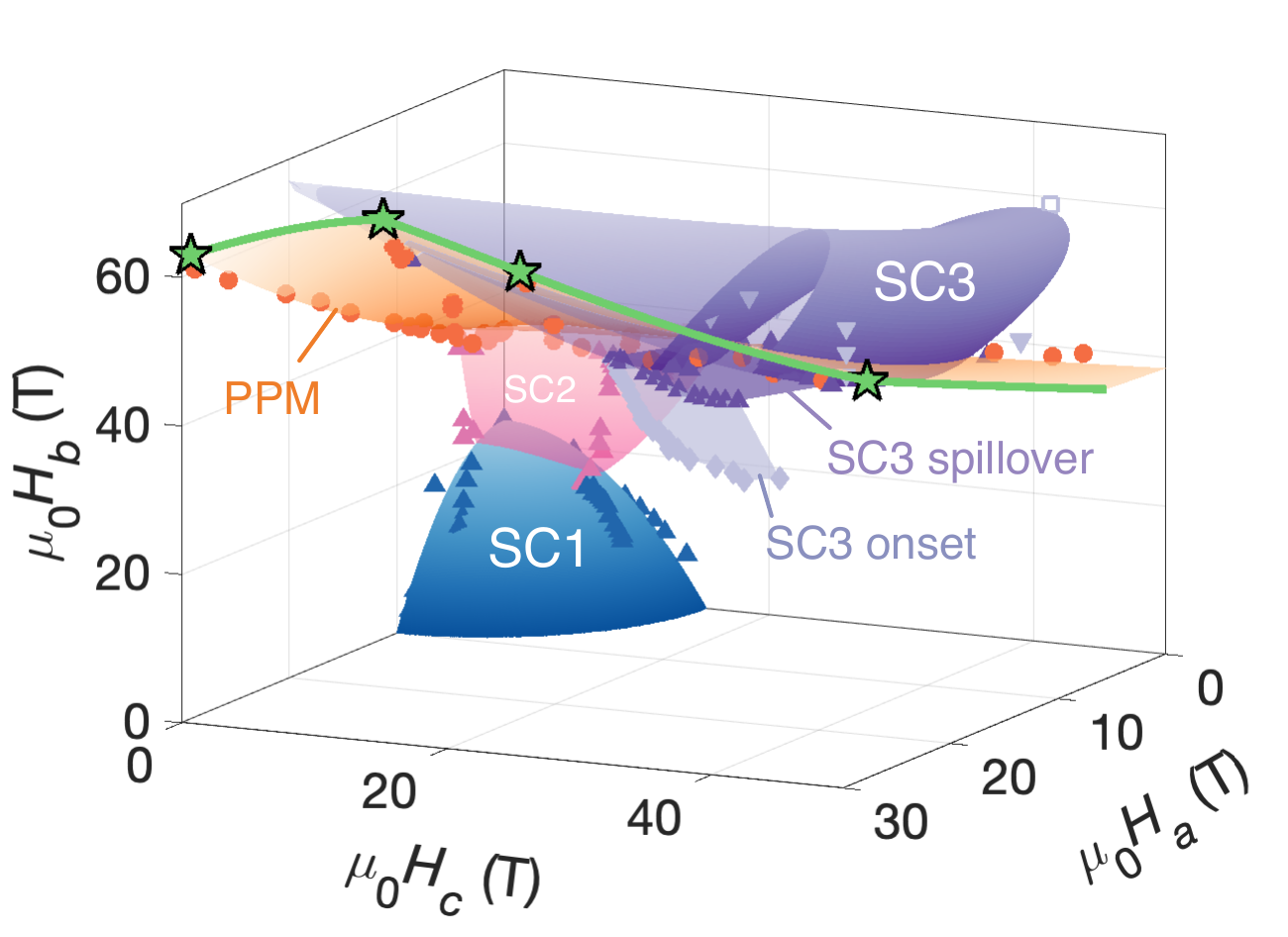} 
    \end{center}
    \vspace{-5pt}
    \caption{Construction of the high magnetic field phase diagram. Here we plot the positive magnetic field octant of the schematic phase diagram displayed in Fig.~\ref{fig:3D_PD}a, including data points used to bound the SC1, SC2, SC3 and PPM phases. Each data point corresponds to either a DC field sweep or a pulsed field measurement. The open square point marking the top of SC3 is taken from ref.~\cite{helm2024}; all other points are from this study. The SC3 spillover and onset regions from Fig.~\ref{fig:SC3_Spill} are also identified.}
    \label{pic:3D_data}
\end{figure}

Our discovery of the presence of an extended line of quantum criticality at high magnetic fields in UTe$_2$ provides a natural explanation for how this material is able to manifest remarkably high-$H$-induced superconductivity. However, understanding the exact shape of the SC3 pocket in \textbf{H}-space remains an outstanding puzzle. In Fig.~\ref{fig:3D_PD}b we schematically depicted the SC3 domain with purple shading, guided by measurements from this study combined with those reported in refs.~\cite{helm2024,lewin2024halo}. (See Figs.~\ref{fig:3D_PD}b \& \ref{pic:3D_data} for the data that constrain this shading, and Fig.~\ref{pic:lewin-comp} for a comparison of our dataset with that of ref.~\cite{lewin2024halo}.) While for $H_a \neq$~0 the QCL closely bounds the highest-$H$ locations of SC3 observed by our measurements and those of ref.~\cite{lewin2024halo}, for $H_a = 0$ (i.e. in the $b-c$ plane) there is a large discrepancy between the termination of the SC3 state and the QCL. In Fig.~\ref{fig:bc-extrpol} we show how the diminution of the MM transition in the $b-c$ plane extrapolates to indicate the QCEP is located at $\mu_0 H \approx$~77~T in this plane, which would correspond to an angle $\theta_{b-c} >$~60$\degree$ (Fig.~\ref{fig:phase-diag-2d}). However, multiple studies by independent groups indicate that the maximal upper critical field and $T_c$ of SC3 in the $b-c$ plane lies in the range 30$\degree<\theta_{b-c}<40\degree$, and appears to terminate at $\theta_{b-c} \approx 45\degree$ (see Fig.~\ref{fig:SC3comparison})~\cite{helm2024,frank2024orphan,tony2025brief}. This is despite the MM transition still giving a sharp discontinuity in $\Delta f_{\text{PDO}}$ at $\theta_{b-c} = 60\degree$ (Fig.~\ref{fig:bc-extrpol}).

This hints at further subtleties in the exotic high-$H$ physical properties of UTe$_2$. One possible explanation concerns the \textbf{d}-vector of SC3 (if it is odd-parity in character, which appears highly likely). When Cooper pairs possess non-zero spin, the gap structure of the superconducting order parameter can set strong constraints on their preferred orientation with respect to \textbf{H}, which may induce complex magnetic field strength-angle-temperature phase diagrams~\cite{TailleferUPt3RevModPhys.74.235,APMackenzieRevModPhys.75.657}. It has recently been proposed that the toroidal shape of the SC3 pocket might be explained by a multi-component non-unitary odd-parity order parameter, with the Cooper pairs possessing a finite orbital momentum predominantly oriented along the $c$-axis~\cite{lewin2024halo}. Given that $c$ is the hard magnetic axis of UTe$_2$ at high $H$~\cite{Miyake2021}, such a scenario appears plausible. Alternatively, perhaps there is no QCEP for $H_a = 0$ (i.e. there should be no open green stars in Fig.~\ref{fig:3D_PD}), but instead the QCL asymptotically bends towards the $H_a = 0$ axis as $H_c \rightarrow \infty$ (i.e. as $\theta_{b-c} \rightarrow 90\degree$). If true, such a scenario would again be analogous to the case of URhGe, in which $a$ is the hard axis and thus $H_a$ has a negligible effect on the $H$-induced superconductivity, which would presumably persist up to arbitrarily large values of $H_a$ until the orbital depairing threshold is reached. Further measurements of UTe$_2$'s SC3 state in proximity to the QCL (at $H_a \neq 0$) by bulk-sensitive thermodynamic probes to track departures from non-Fermi liquid phenomenology, and by NMR of the spin relaxation rates upon approaching the QCL -- although challenging due to the high magnetic field scales and practical difficulties in sample orientation -- would be valuable for shedding further light onto the microscopic properties of the remarkable three-dimensional quantum critical phase boundary manifested by this material.

In summary, here we have presented a detailed mapping of the high magnetic field phase landscape of UTe$_2$. We tracked the profile of this material's $H$-induced first-order metamagnetic transition surface as functions of temperature and of magnetic field tilt angle. We find that for rotations of \textbf{H} away from the crystallographic $b$-axis, the critical end point of the metamagnetic transition is smoothly suppressed towards zero temperature. Because the metamagnetic transition surface extends throughout the three dimensions of magnetic field space, this implies that it is bounded by a continuous ring of quantum critical end points, the locus of which forms an extended line of quantum criticality. For \textbf{H} oriented close to this quantum critical line, the high-$H$ SC3 superconductive state is observed on both sides of the metamagnetic transition, similar to the case of $H$-induced superconductivity in the related material URhGe~\cite{levy2005magnetic,levy2007acute}. Magnetic fluctuations emanating from this three-dimensional quantum critical phase boundary therefore appear the likely source of electronic pairing mediating the formation of $H$-induced superconductivity in UTe$_2$.

\begin{figure*}[h!]
\vspace{-3cm}
\begin{center}
\includegraphics[width=\linewidth]{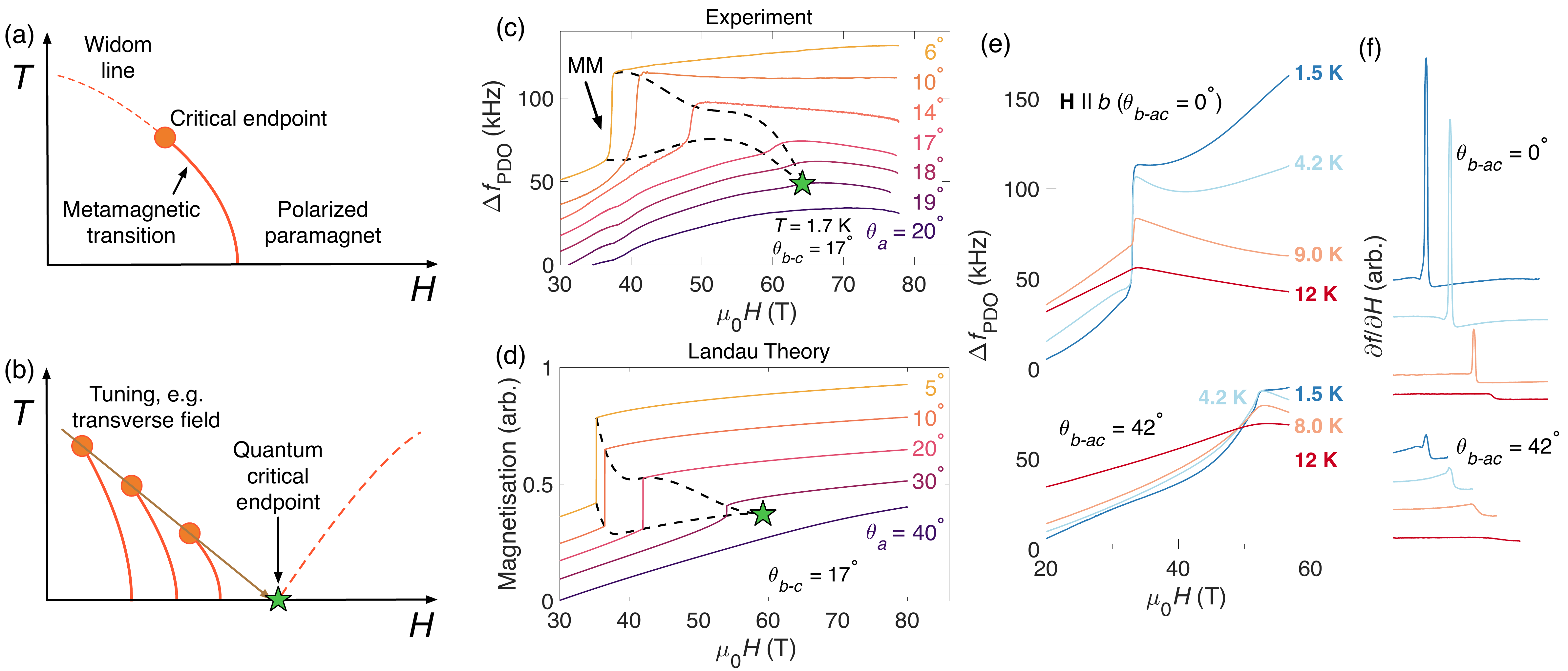}
\end{center}
\caption{Suppression of a first-order metamagnetic transition towards zero temperature. (a) Schematic temperature $T$ versus magnetic field $H$ depiction of the critical endpoint (CEP) of a first-order metamagnetic transition. The solid orange line marks the first-order phase boundary, while the dashed (Widom) line indicates crossover behavior at higher temperatures. (b) The modulation of a non-thermal tuning parameter -- such as a magnetic field -- can suppress a CEP to lower temperatures. When it reaches absolute zero, it becomes a quantum critical endpoint (QCEP). (c) Isothermal contactless resistivity measurements of UTe$_2$, obtained by the PDO technique (see Appendix~\ref{appx_exp-tech}), offset for clarity. The sample was mounted at a fixed angle of 17$\degree$ from $b$ towards $c$, and subsequently rotated towards $a$ (see Fig.~\ref{fig:wedge-sketch} for a description of this rotation procedure). At a tilt of 6$\degree$ away from the $bc$ plane the metamagnetic (MM) transition is sharp and first-order-like. However, at $\theta_a$~=~18$\degree$ it has markedly softened and by 20$\degree$ the QCEP has been passed. Dashed lines and green star give a visual depiction of how the tilt angle of the magnetic field tunes the CEP to a QCEP. (d) Simulation of magnetization for rotations in the same plane as panel (c), calculated from Eqn.~\ref{F_metamagnetic} showing how the sharp jump in magnetization is suppressed upon approaching a QCEP. (e) Temperature evolution of the MM transition for \textbf{H}~$\parallel b$ (top) and at $\theta_{b-ac} =$~42$\degree$ (bottom); offset derivatives are plotted in panel (f). The 9.0~K transition for \textbf{H}~$\parallel b$ is considerably sharper than the 1.5~K curve at $\theta_{b-ac}$~=~42$\degree$, indicating that the CEP is located at a much lower temperature at this angle}.
\label{fig:MMdeath}
\end{figure*}

\begin{figure}[t!]
    \vspace{0cm}
     \begin{center}
     \includegraphics[width=0.9\linewidth]{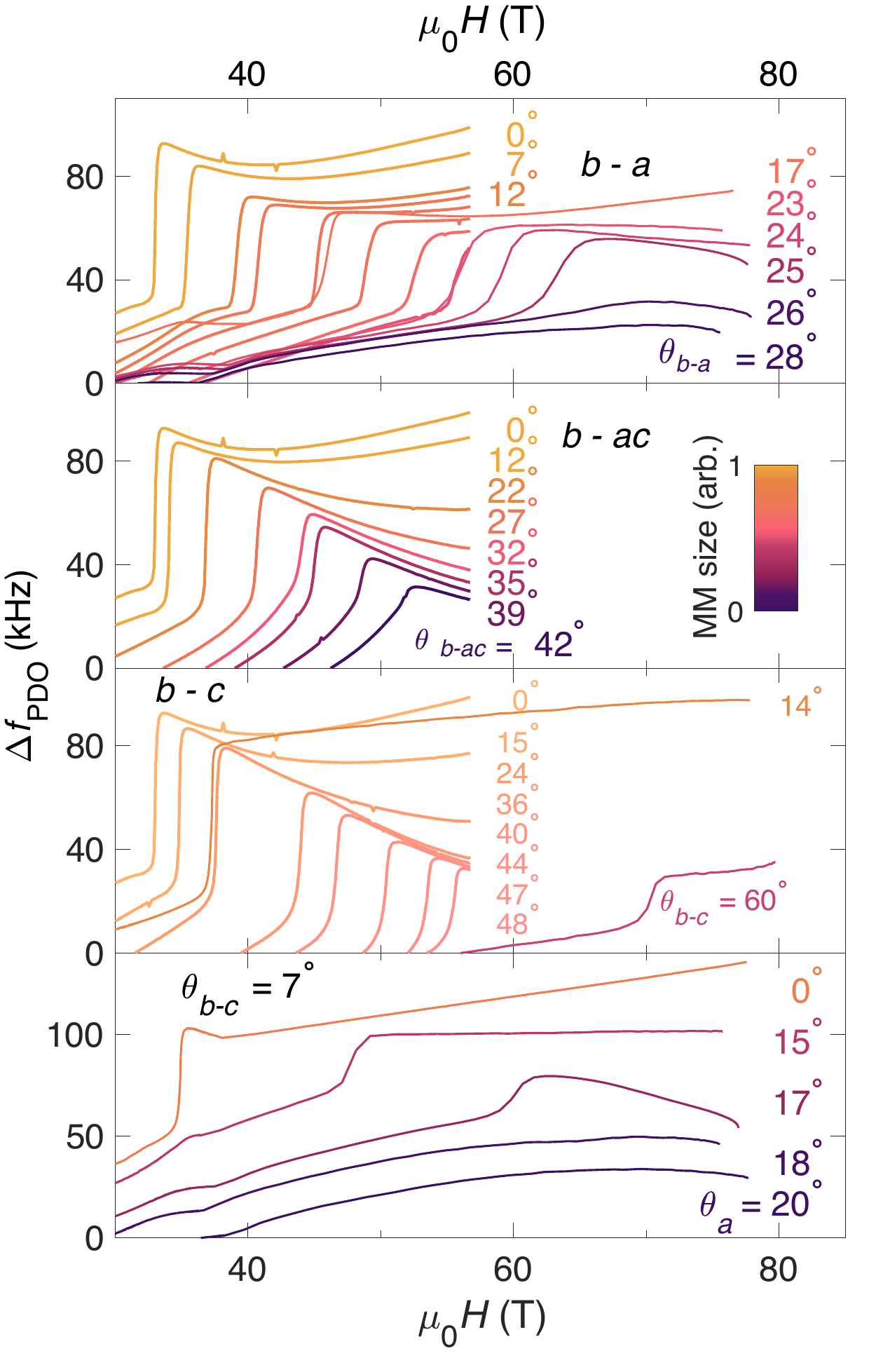} 
     \end{center}
     \vspace{-5pt}
     \caption{Mapping quantum criticality in three dimensions. PDO data for the $b-a$ ([010]--[100]), $b-ac$ ([010]--[101]) and $b-c$ ([010]--[001]) rotation planes, along with rotations towards $a$ at a fixed inclination of 7$\degree$ from $b$ to $c$. All curves to $<$~60~T were measured in a single coil magnet in Wuhan at $T =$~4.2~K, while those to $>$~60~T were measured in a dual-coil magnet in Dresden at 1.7~K; note that the Dresden data have been rescaled for ease of comparison, while the nature of the dual-coil magnet system introduces a small artifact at 39~T (see Fig.~\ref{pic:coil-times} for details). For each of these rotation planes, the size of the MM transition decreases as \textbf{H} rotates further from $b$, as expected for the behavior depicted in Fig.~\ref{fig:MMdeath}b. This is especially pronounced in the bottom panel, where a small change in inclination from 17$\degree$ to 18$\degree$ abruptly crosses the CEP at this relatively low measurement temperature of 1.7~K.}
    \label{fig:4panel-MM}
\end{figure}

\clearpage

\begin{figure}[t!]
\vspace{0cm}
\begin{center}
\includegraphics[width=1\linewidth]{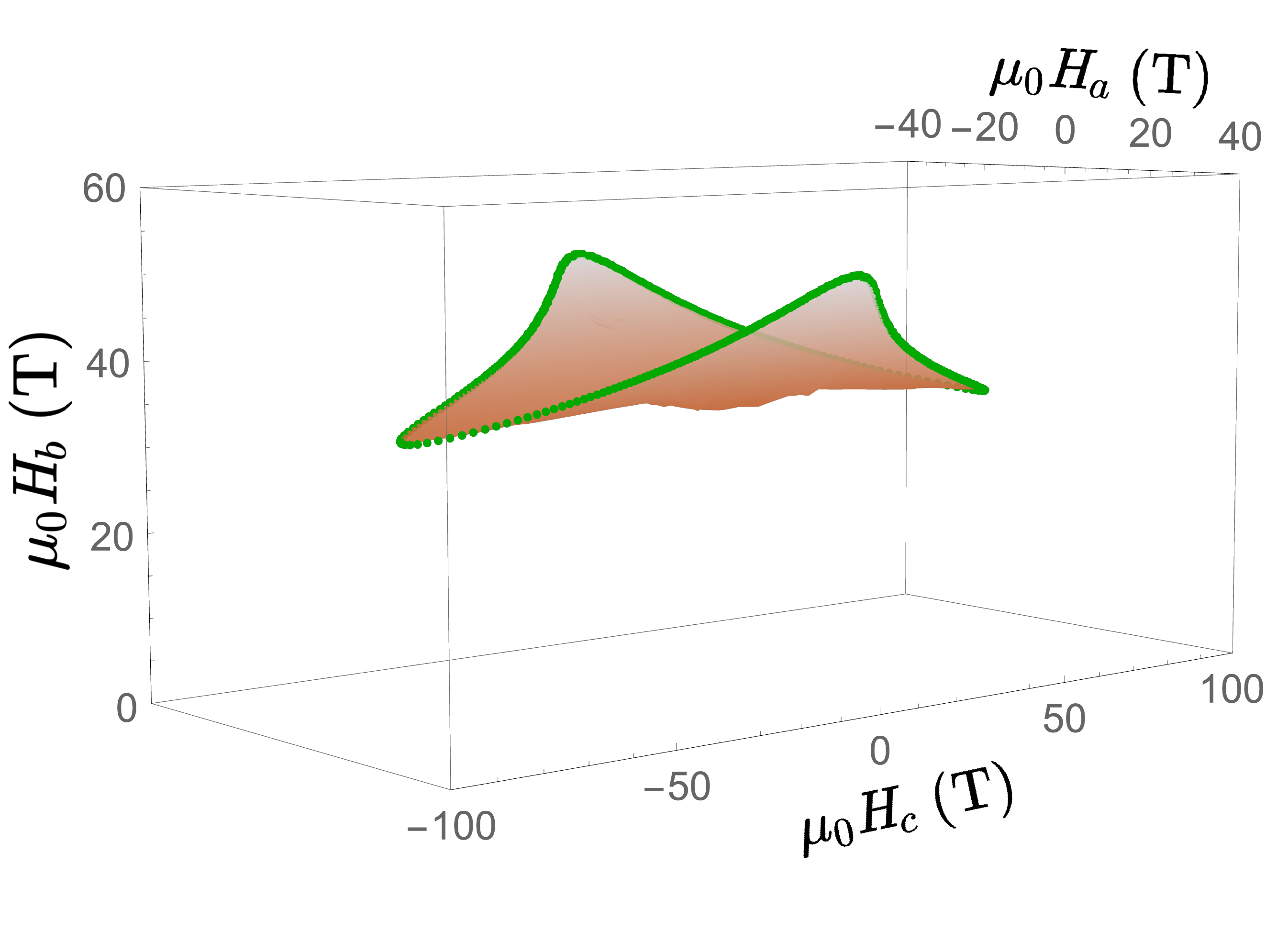}
\end{center}
\caption{Numerical simulation of the MM transition and QCL using a Landau free energy expansion. As in Fig. \ref{fig:3D_PD}, the orange surface corresponds to the surface in \(\mathbf{H}\)--space at which the first-order MM transition occurs, obtained by interpolation from numerically determined locations of the jumps in the magnetization found by minimizing the free energy of Eqn.~\eqref{F_metamagnetic}. The QCL (green points) was obtained by determining the location where the jumps vanish within numerical accuracy. The parameters were chosen as follows: \(\chi_{a}^{-1}=36.13\) (easy axis), \(\chi_{b}^{-1}=\beta_{cc}=180.7\) (hard axis), \(\chi_{c}^{-1}=\beta_{aa}=72.27\), \(\beta_{bb}=-688.4\), \(\beta_{ab}=0.18\), \(\beta_{bc}=-361.3\), \(\delta_{bbb}=903.4\), \(\delta_{ccc}=1445\), \(\delta_{abb}=903.4\), \(\delta_{bbc}=1084\), and all others zero.
}
\label{fig:QCLsim}
\end{figure}


\begin{acknowledgments}
    We are grateful to A. Levchenko, N.R. Cooper, T. Helm, M. Kwasigroch, A. Huxley, T. Hazra, P. Coleman and J. Schmalian for stimulating discussions. We thank S.T. Hannahs for technical advice and assistance. This project was supported by the EPSRC of the UK (grant no. EP/X011992/1). A portion of this work was performed at the National High Magnetic Field Laboratory, which is supported by National Science Foundation Cooperative Agreement Nos. DMR-1644779 \& DMR-2128556 and the State of Florida. We acknowledge support of the HLD at HZDR, a member of the European Magnetic Field Laboratory (EMFL). The EMFL also supported dual-access to facilities at MGML, Charles University, Prague, under the European Union's Horizon 2020 research and innovation programme through the ISABEL project (No. 871106). Crystal growth and characterization were performed in MGML (mgml.eu), which is supported within the program of Czech Research Infrastructures (project no. LM2023065). We acknowledge financial support by the Czech Science Foundation (GACR), project No. 22-22322S. T.I.W. and A.J.H. acknowledge support from EPSRC studentships EP/R513180/1 \& EP/M506485/1. T.I.W. and A.G.E. acknowledge support from ICAM through US National Science Foundation (NSF) Grant Number 2201516 under the Accelnet program of the Office of International Science and Engineering and from QuantEmX grants from ICAM and the Gordon and Betty Moore Foundation through Grants GBMF5305 \& GBMF9616. D.V.C. acknowledges financial support from the National High Magnetic Field Laboratory through a Dirac Fellowship, which is funded by the National Science Foundation (Grant No. DMR-1644779) and the State of Florida. The work at UW-Madison (D.S.) was financially supported by the National Science Foundation, Quantum Leap Challenge Institute for Hybrid Quantum Architectures and Networks Grant No. OMA-2016136.
A.G.E. acknowledges support from the Henry Royce Institute for Advanced Materials through the Equipment Access Scheme enabling access to the Advanced Materials Characterisation Suite at Cambridge, grant numbers EP/P024947/1, EP/M000524/1 \& EP/R00661X/1; and from Sidney Sussex College (University of Cambridge).
\end{acknowledgments}
\clearpage

\appendix
\section{EXPERIMENTAL TECHNIQUES}
\label{appx_exp-tech}

\normalsize
\vspace{5mm}\noindent 
\textbf{Sample preparation}\\ High quality single crystals of UTe$_2$ were grown by the molten salt flux method, adapted from ref.~\cite{PhysRevMaterials.6.073401MSF_UTe2}. Samples were grown by the methodology given in ref.~\cite{Eaton2024}. Specimens were oriented by single crystal Laue diffractometry, and screened by a combination of residual resistivity, specific heat, and magnetization measurements.

\begin{figure}[h!]
    \vspace{0cm}
     \begin{center}
     \includegraphics[width=.9\linewidth]{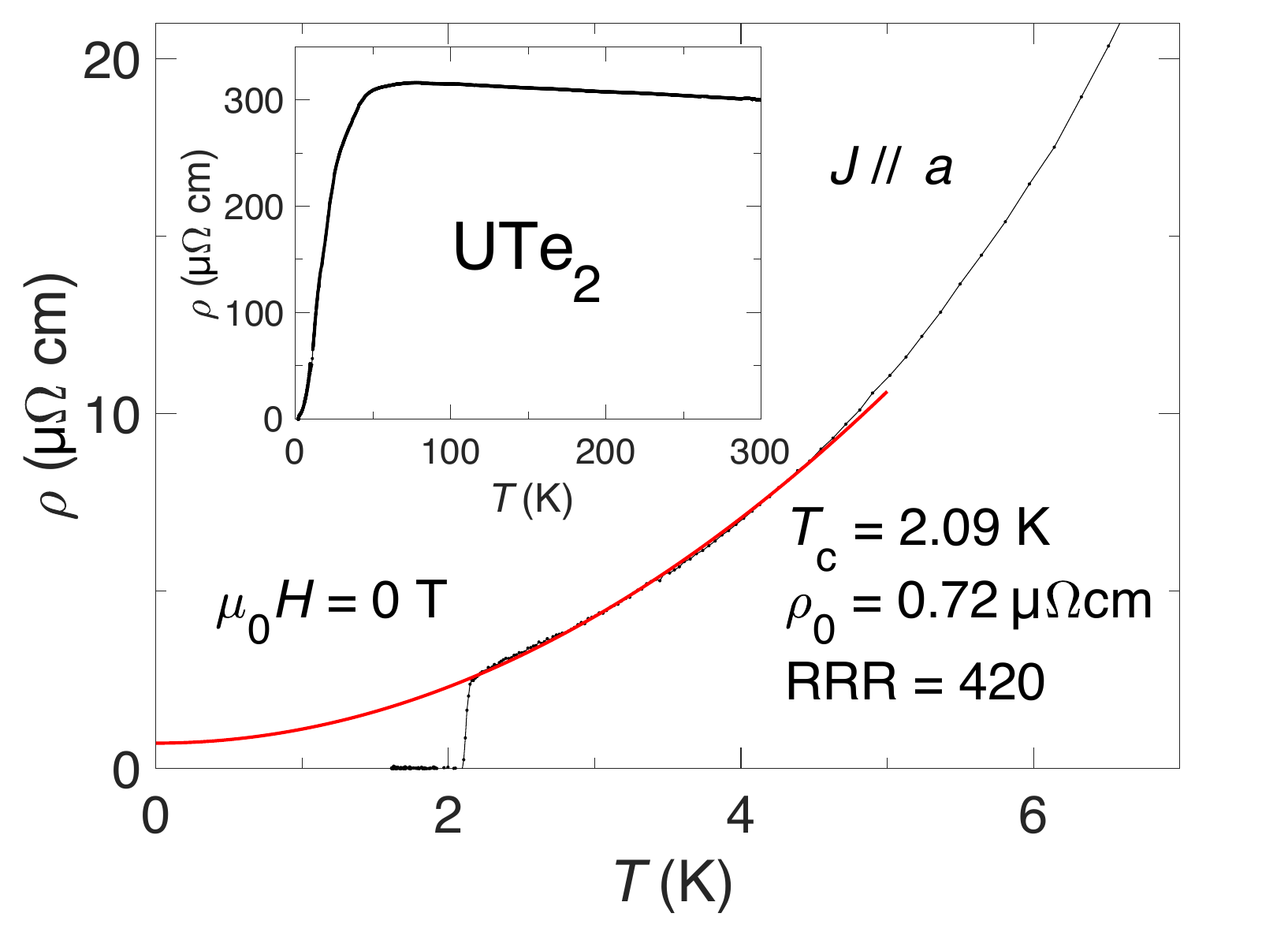} 
     \end{center}
     \vspace{-5pt}
     \caption{Zero field contacted resistivity versus temperature measured by the four-terminal technique on the same sample as that investigated in DC magnetic fields up to 41.5~T for which data are presented in Figure~\ref{fig:SC3_Spill}. Current density $J$ was sourced parallel to the $a$-axis. The $T_{\text{c}}$ value of 2.09~K is determined from the resistivity falling to zero (within the resolution of the measurement). Extrapolating the normal state resistivity below 5~K with a quadratic fitting towards zero temperature (red curve) yields a residual resistivity ratio (RRR) of 420, indicative of high crystalline quality~\cite{PhysRevMaterials.6.073401MSF_UTe2,tony2024enhanced,aoki2024molten}.}
    \label{fig:RRR}
\end{figure}

\vspace{5mm}\noindent 
\textbf{Electrical conductivity measurements}\\ Contacted transport measurements were performed by the four-wire technique using low frequency AC excitations ($<$ 50 Hz) with voltages probed by phase sensitive detection. Electrical contacts were made by spot-welding gold wires onto samples, which were then mechanically secured with low-temperature epoxy.

\begin{figure}[h!]
    \vspace{0cm}
     \begin{center}
     \includegraphics[width=1\linewidth]{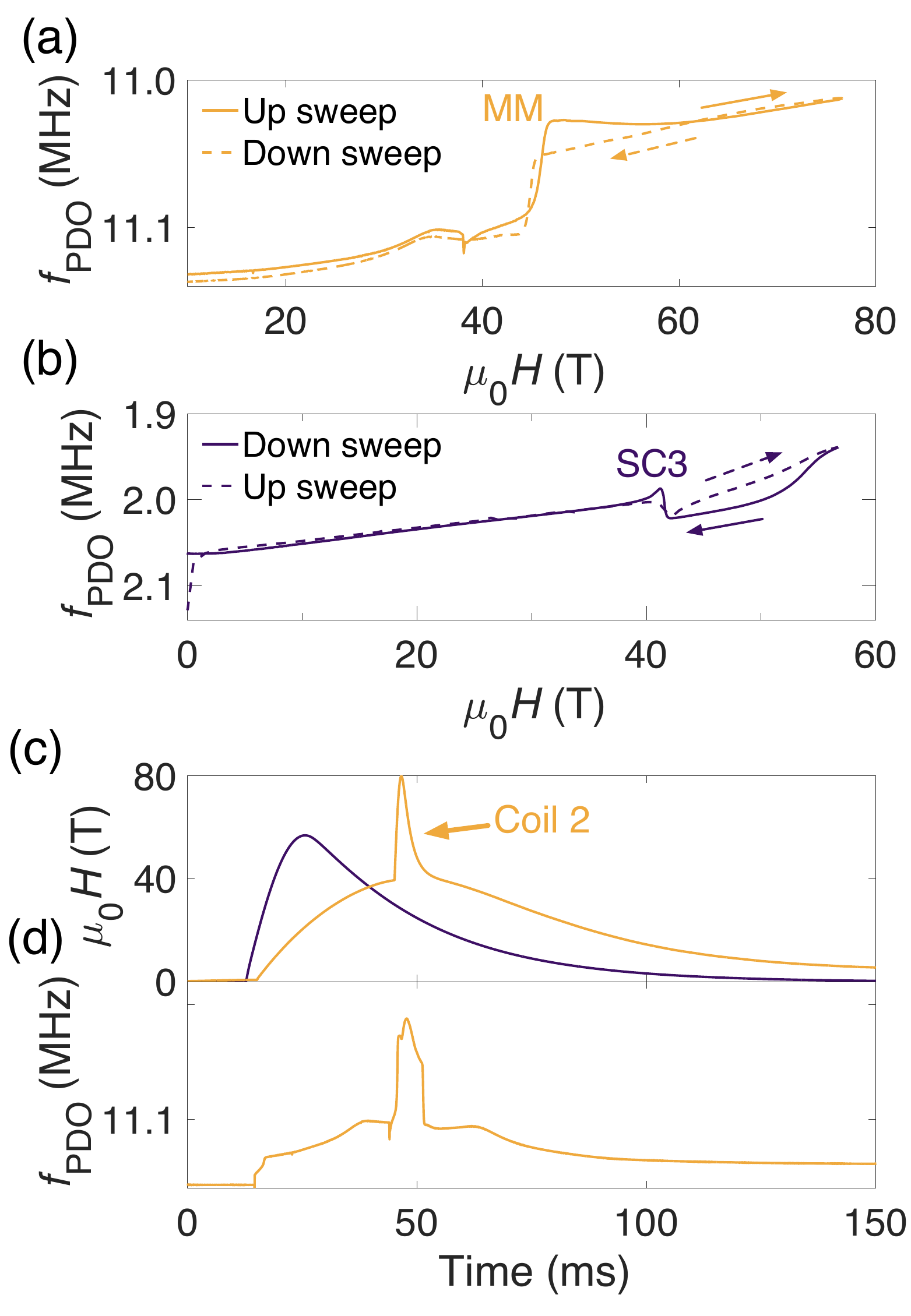} 
     \end{center}
     \vspace{-5pt}
     \caption{(a) Raw PDO data from a pulsed field sweep using the Dresden dual-coil magnet system. (b) Raw PDO data from a pulsed field sweep using single-coil magnet system in Wuhan. (c) Magnetic field versus time for representative magnetic field pulses performed in Dresden (indigo) and Dresden (ochre). An arrow indicates the field profile of the second coil in Dresden. (d) The same data as in (a) plotted as a function of time. The small artifacts in our Dresden measurements at 39~T stem from the activation of the second coil, and are not intrinsic features of UTe$_2$, unlike the metamagnetic transition.}
    \label{pic:coil-times}
\end{figure}

Contactless electrical conductivity measurements were made by the proximity detector oscillator (PDO) technique~\cite{PDO_Altarawneh}, following the same methodology as our prior UTe$_2$ PDO study~\cite{theo2024}. Experiments were performed at the Wuhan National High Magnetic Field Center in pulsed magnetic fields up to 56.7~T, and at the Hochfeld-Magnetlabor Dresden (HLD-EMFL) in a dual-coil magnet to a maximum applied field of 79.7~T. PDO measurements were performed at frequencies of tens of MHz, giving good resolution despite the short duration of a magnetic field pulse, of order 0.1~s (see Figure~\ref{pic:coil-times}). By tracking the change in frequency $f$ of the PDO circuit~\cite{PDO_Altarawneh,ghannadzadeh2011PDO} as a function of $H$, the change in a sample's electrical resistivity $\rho$ is captured as

\begin{equation} \label{Eq:dff}
    \frac{\Delta f}{f} \approx -\eta\frac{\delta}{d} \left(\mu_r \frac{\Delta \rho}{\rho}+\Delta \chi_s \right),
\end{equation}

\noindent
for a sample of thickness $d$, where $\mu_r = \chi_s+1$ and the skin depth $\delta = \sqrt{\frac{2\rho}{\mu_r\mu_0\omega}}$, where $\omega$ is the excitation frequency~\cite{PDO_Altarawneh,ghannadzadeh2011PDO}. Our PDO circuitry typically rings in the range 28-32~MHz -- in Dresden this signal was then mixed down once to $\approx$~11~MHz and acquired using a Teledyne LeCroye HDO6000B oscilloscope acquiring at 500 megasamples per second. In Wuhan we mixed down twice and acquired at $\approx$~2~MHz utilizing a JYTEK PXIe-69834 data acquisition card at a rate of 30 megasamples per second.


\begin{figure}[h!]
    \vspace{0cm}
     \begin{center}
     \includegraphics[width=.7\linewidth]{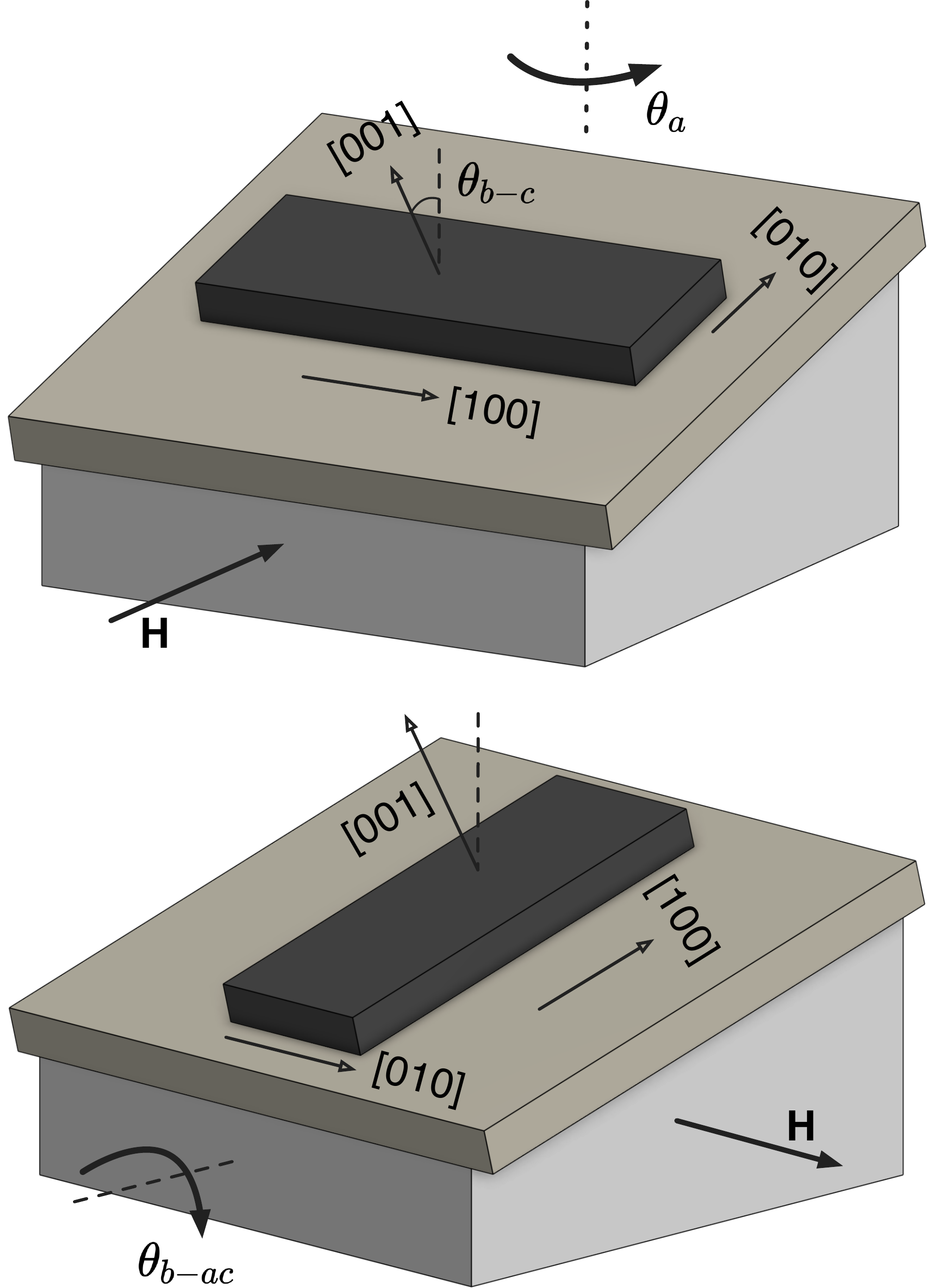} 
     \end{center}
     \vspace{-5pt}
     \caption{Schematic representation of the experimental setup used for simultaneously applying $H_a$, $H_b$ and $H_c$ magnetic field components to UTe$_2$. Samples were mounted on G10 wedges, giving a fixed inclination between the $c$-axis and either $a$ or $b$. By rotating the orientation of the magnetic field about an axis orthogonal to the axial inclination of the wedge, the applied components of $H_a$, $H_b$ and $H_c$ were modulated.}
    \label{fig:wedge-sketch}
\end{figure}

\section{LANDAU THEORY OF METAMAGNETIC TRANSITION AND QUANTUM CRITICAL LINE}
\label{app-theory}
\noindent
The theoretical simulation of magnetization curves in Fig. \ref{fig:MMdeath}b was obtained by numerically searching for local minima of the free energy in Eq.~\ref{F_metamagnetic}, picking the lowest minimum found. The critical field \(\mathbf{H}_m(\theta_{b-c},\theta_{a-c})\) at which the MM transition occurs for a fixed field direction was identified by iteratively searching for the largest rate of change in the magnetization. The QCEP magnetic field values \(\mathbf{H}_{QCEP}(\theta_{a-c})\) were then obtained by scanning over different field directions starting from the \(b\) axis and incrementing the \(\theta_{b-c}\) and \(\theta_{a-c}\) angles in steps of \(1.8\degree\), terminating when the numerically computed slope of the magnetization 
curve along the scanning direction at the MM transition fell below a threshold (determined by the magnetization slopes above and below the transition). The locus of the points \(\mathbf{H}_m(\theta_{b-c},\theta_{a-c})\) and \(\mathbf{H}_{QCEP}(\theta_{a-c})\) thus obtained were used, respectively, to plot the MM phase transition surface (orange, obtained by interpolation) and the QCL line (green) in Figure~\ref{fig:QCLsim}.

Now we turn to provide more qualitative details of the Landau theory of the metamagnetic (MM) phase transition described by the free energy in Eq.~\ref{F_metamagnetic} in the main text:
\begin{align}
    \mathcal{F}[\mathbf{M}](\mathbf{H})&=\frac{1}{2}\chi^{-1}_{i}M_i^2+\frac{1}{4}\beta_{ij} M_i^2M_j^2\\&+ \frac{1}{6} \delta_{ijk} M_i^2 M_j^2 M_k^2 -\mathbf{M\cdot H}.
\label{F_metamagnetic-app}
\end{align}
To understand the origin of the critical end point (CEP) and the quantum critical end point (QCEP), we first review a simpler case of a 1D model:
\begin{align}
    \mathcal{F}[M](H)&=\frac{1}{2}\chi^{-1}M^2+\frac{1}{4}\beta M^4\\&+ \frac{1}{6} \delta M^6 - MH.
\label{F_metamagnetic1D}
\end{align}
The ground state is found by minimizing the free energy with respect to \(M\), which gives
\[H=\chi^{-1} M + \beta M^3 + \delta M^5.\]
At the MM transition, \(\partial_M H=0\), which can be solved analytically for \(M\):
\[M^2_\pm=\frac{-3 \beta \pm \sqrt{9 \beta^2 - 20 \chi^{-1} \delta}}{10 \delta}\]
with \(M_\pm\) being the values of \(M\) immediately above and below the MM transition, respectively. In order for the MM transition to exist, we must therefore have \(\beta<0\) and \(9 \beta^2 - 20 \chi^{-1} \delta>0\), and the MM jump vanishes precisely when \(9 \beta^2 - 20 \chi^{-1} \delta=0\). The latter condition thus determines the CEP when the Landau parameters \(\chi^{-1}, \beta,\) and \(\delta\) are functions of some external parameter like temperature \(T\):
\[9 \beta^2(T_{CEP}) - 20 \chi^{-1}(T_{CEP}) \delta(T_{CEP})=0.\]
In the case of UTe$_2$, $T_{CEP}\approx$~10~K for \textbf{H}~$\parallel b$~\cite{Miyake2019,valiska2024dramatic}.

Generically, the CEP occurs due to \(\beta(T)\) increasing with increasing temperature and eventually changing sign at some temperature \(T_0>T_{CEP}\). As an example, in itinerant systems, to leading order close to temperature \(T_0\) at which \(\beta(T)=0\), the temperature dependence of the coefficients can be modelled phenomenologically as \cite{Yamada_metamagnetic}
\begin{align}
     \chi^{-1}_i(T)&\propto 1 + b T^2 - bc T^4\nonumber,\\
     \beta_{ij}(T)&\propto 1 - cT^2.
\end{align}
When the parameters \(b\) and \(c\) are 
functions of another external parameter \(\theta\), i.e. the field direction, the corresponding \(T_{CEP}\) also becomes a function of \(\theta\) and may even vanish for some \(\theta_{QCP}\), resulting in a quantum critical point.

To obtain a qualitative understanding of the 3D free energy in Eq. (\ref{F_metamagnetic}), we can rewrite \(\mathbf{M}\) in spherical coordinates, which effectively gives
\begin{align}
    \mathcal{F}[M,\theta,\phi](\mathbf{H})&=\frac{1}{2}\chi^{-1}(\theta,\phi)M^2+\frac{1}{4}\beta(\theta,\phi) M^4\\ &+ \frac{1}{6} \delta(\theta,\phi) M^6\nonumber
    -MH(\sin\phi\sin\phi_H\\&\cdot\cos(\theta-\theta_H)+\cos\phi\cos\phi_H)
\label{F_metamagneticSpherical}
\end{align}
where \(\theta\) and \(\phi\) are azimuthal and polar angles of \(\mathbf{M}\) and \(\theta_H\) and \(\phi_H\) are azimuthal and polar angles of \(\mathbf{H}\). For fixed \(\mathbf{H}\), minimizing over \(\theta\) and \(\phi\) determines the direction of \(\mathbf{M}\), which implicitly determines \(\theta(\mathbf{H})\) and \(\phi(\mathbf{H})\). Consequently, we can consider \(\chi^{-1}, \beta\), and \(\delta\) to be functions of \(\mathbf{H}\), as well as temperature. Strictly speaking, we should add a coefficient in front of the \(MH\) term in Eq. \ref{F_metamagneticSpherical} that is also a function of \(\mathbf{H}\), adding a nonlinearity in \(H\) to the problem; however, since the CEP condition is independent of \(H\) we can neglect this nonlinearity as long as it is not too strong. Setting the temperature to zero, we then have a QCP condition given by
\[9 \beta^2(\mathbf{H}_{*}) - 20 \chi^{-1}(\mathbf{H}_{*}) \delta(\mathbf{H}_{*})=0.\]
This equation defines a 2D surface in the 3D space of \(\mathbf{H}\), which generically intersects the 2D MM transition surface (determined by \(\partial_\mathbf{M}\mathbf{H}=0\)) in a line, thereby forming the quantum critical line (QCL). Qualitatively, given that \(\beta(H_a,H_b,H_c)\) is minimal for \(H_a=H_c=0\), the field components \(H_a\) and \(H_c\) act effectively in the same way as 
temperature as they increase \(\beta\) towards zero.

\section{SUPPORTING FIGURES}

\begin{figure}[h!]
    \vspace{-0cm}
    \begin{center}
    \includegraphics[width=1\linewidth]{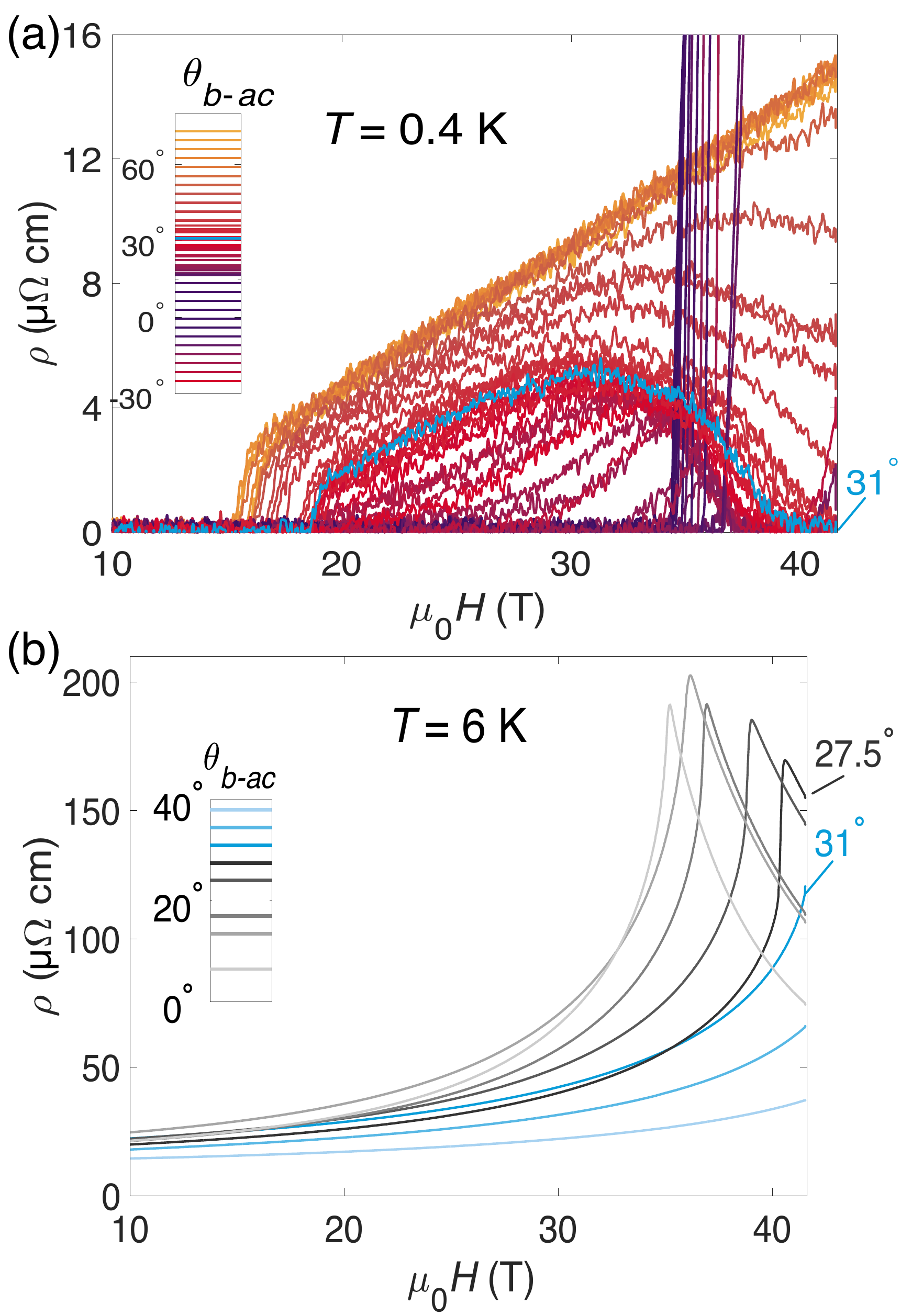} 
    \end{center}
    \vspace{-5pt}
    \caption{Identification of the spillover of SC3 outside the PPM. (a) DC magnetic field sweeps at $T$~=~0.4~K in the $b-ac$ rotation plane with $J =$~150~mA/cm$^2$. A subset of these data are presented in Fig.~\ref{fig:SC3_Spill}a. These data are used to construct the phase diagram in Fig.~\ref{fig:SC3_Spill}b. (b) DC field sweeps at the elevated temperature of $T$=~6~K -- above all critical superconducting temperatures -- in order to map the extent of the PPM phase in this rotation plane. The MM transition is observed as a peak in $\rho(H)$ at incremental rotation angles up to $\theta_{b-ac}$~=~27.5$\degree$; however, at the next measured angle of $\theta_{b-ac}$~=~31.0$\degree$ no peak is observed up to the highest applied field of 41.5~T. In panel (a) the 0.4~K data at $\theta_{b-ac}$~=~31.0$\degree$ is plotted in the same blue as panel (b), clearly showing that the zero resistance state of SC3 is accessed even though the MM transition for this field orientation lies at higher field strengths.}
    \label{fig:blue-spill}
\end{figure}

\begin{figure*}[h!]
    \vspace{-0cm}
    \begin{center}
    \includegraphics[width=.8\linewidth]{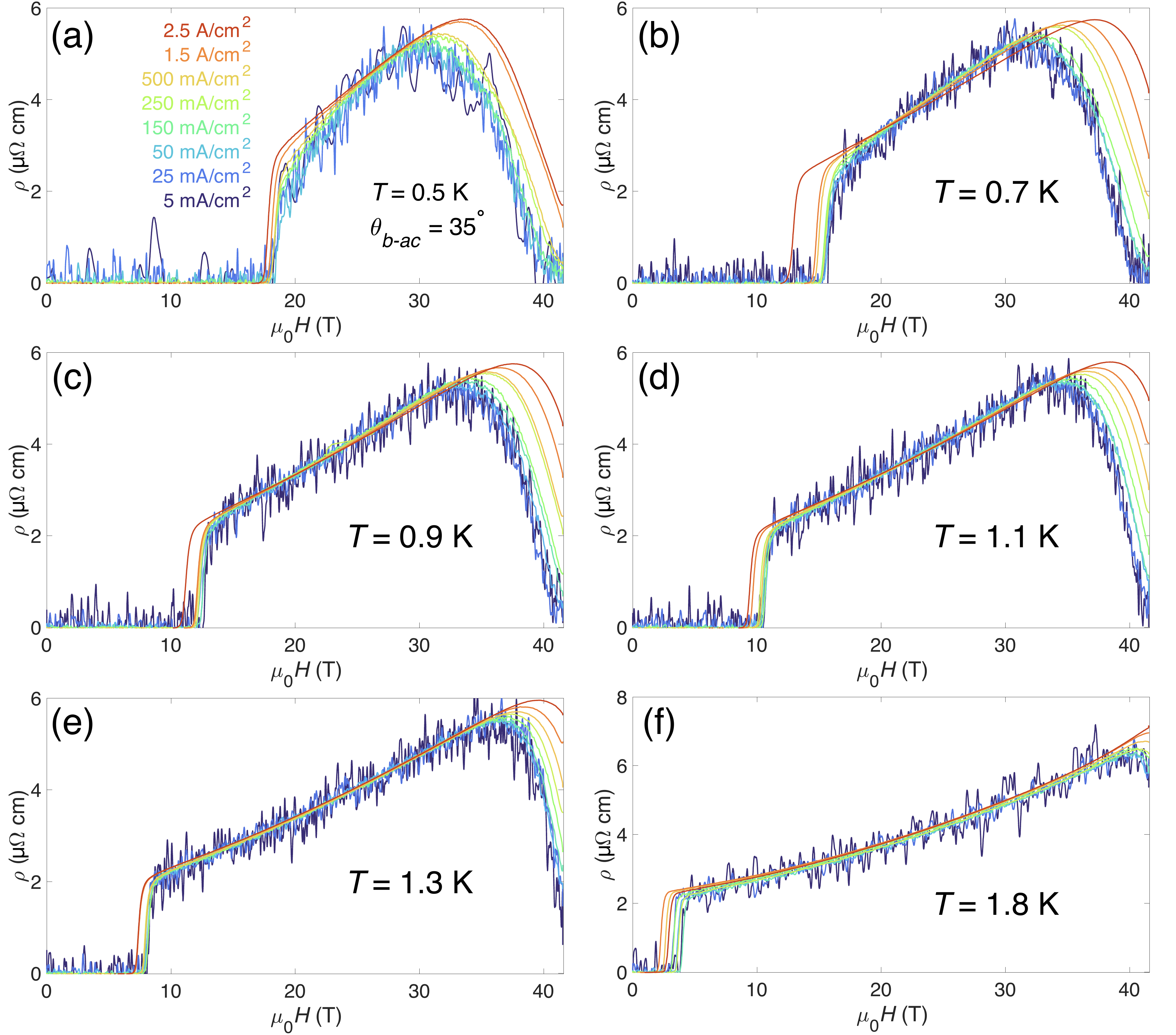} 
    \end{center}
    \vspace{-5pt}
    \caption{Non-Ohmic current--voltage behavior near the QCL. Effective resistivity $\rho$ for incremental currents at isothermal temperature points as indicated. We note that some Joule heating likely occurs for the highest current measurements at 0.5~K (and to a lesser extent for 0.7~K) as seen by the change in slope of $\rho(H)$ in the normal state over the interval 25~T~$\lessapprox \mu_0 H \lessapprox$~30~T for e.g. the 2.5 A/cm$^2$ curve compared to that at 50~mA/cm$^2$. However, there is clearly an intrinsic dependence on current for the onset of SC3 between the 150~mA/cm$^2$ and 500~mA/cm$^2$ curves, which exhibit no indication of extrinsic heating effects. Furthermore, at higher temperatures the linear relationship between $V$ and $J$ in the normal state (i.e. at $\mu_0 H =$~25~T) holds extremely well for all currents, while even up to 1.8~K at $\mu_0 H \approx$~40~T the SC3 onset regime exhibits clearly non-linear $J-V$ behavior. The 1.1~K data are those displayed in Fig.~\ref{fig:SC3_Spill}e.}
    \label{pic:IV_all}
\end{figure*}

\begin{figure}[h!]
    \vspace{-0cm}
    \begin{center}
    \includegraphics[width=1\linewidth]{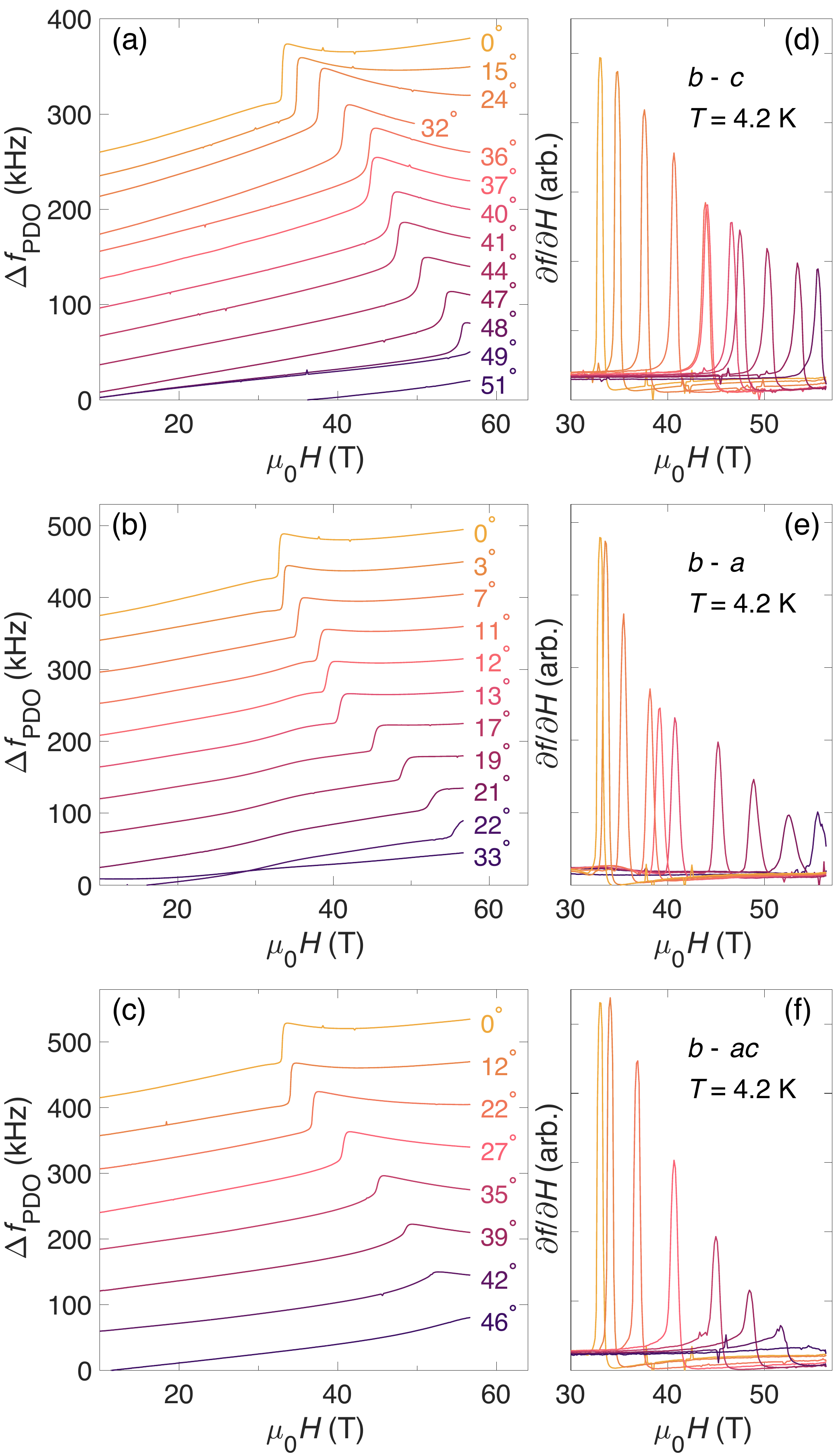} 
    \end{center}
    \vspace{-5pt}
    \caption{Tracking the softening of the metamagnetic transition as the QCL is approached. Contactless resistivity measurements by the PDO technique in the (a) $b-c$, (b) $b-a$ and (c) $b-ac$ rotation planes, with corresponding derivatives in (d-f). All data were collected on the same sample at $T =$~4.2~K, measured in the Wuhan National High Magnetic Field Center.}
    \label{pic:Wuhan_rotation_all}
\end{figure}

\begin{figure}[h!]
    \vspace{-0cm}
    \begin{center}
    \includegraphics[width=1\linewidth]{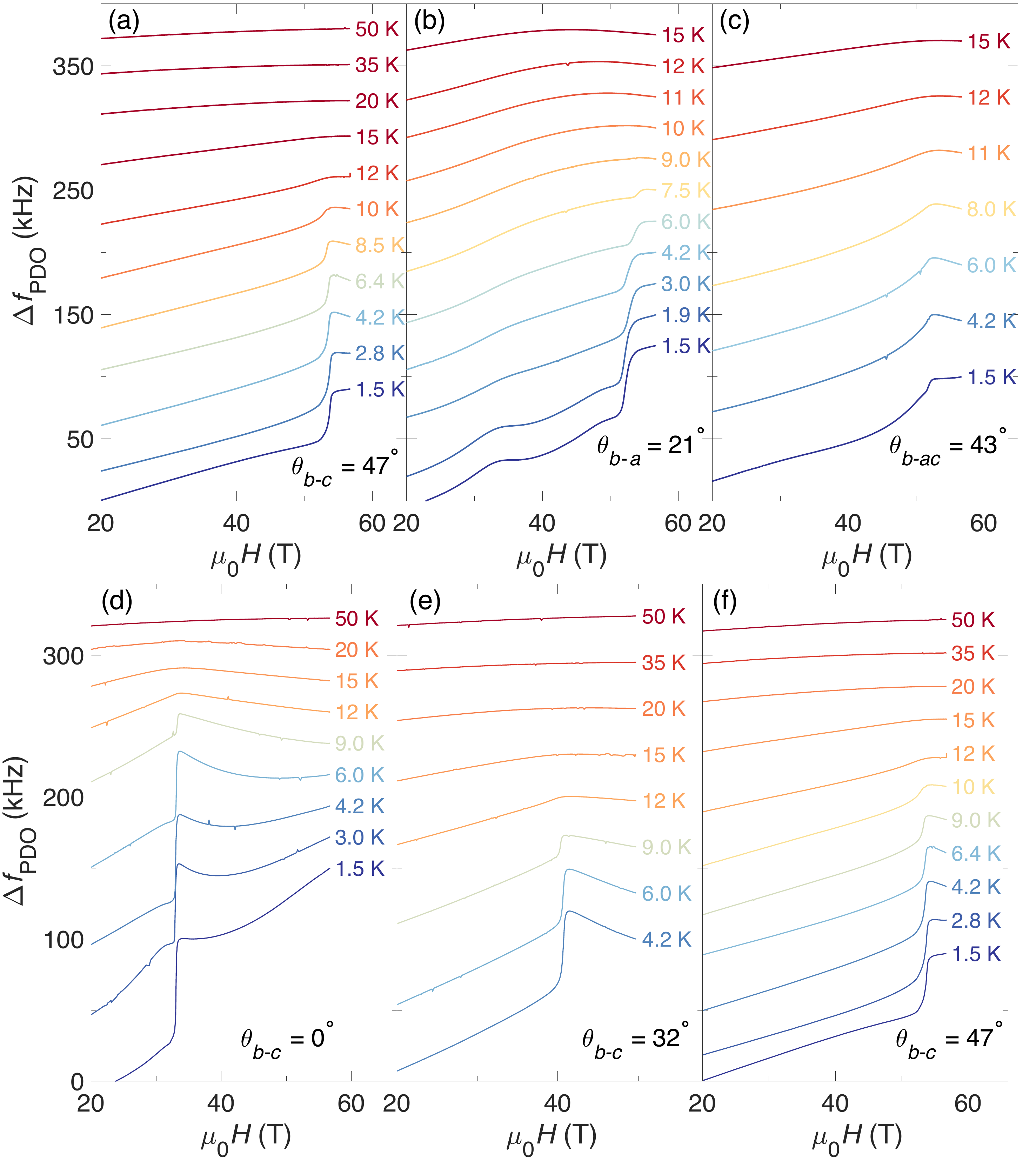} 
    \end{center}
    \vspace{-5pt}
    \caption{Temperature evolution of the metamagnetic transition. PDO data at incremental temperatures for six different magnetic field orientations, as labeled. All data were collected on the same sample, measured in the Wuhan National High Magnetic Field Center.}
    \label{pic:Wuhan_HT_all}
\end{figure}

\begin{figure}[h!]
    \vspace{0cm}
     \begin{center}
     \includegraphics[width=1\linewidth]{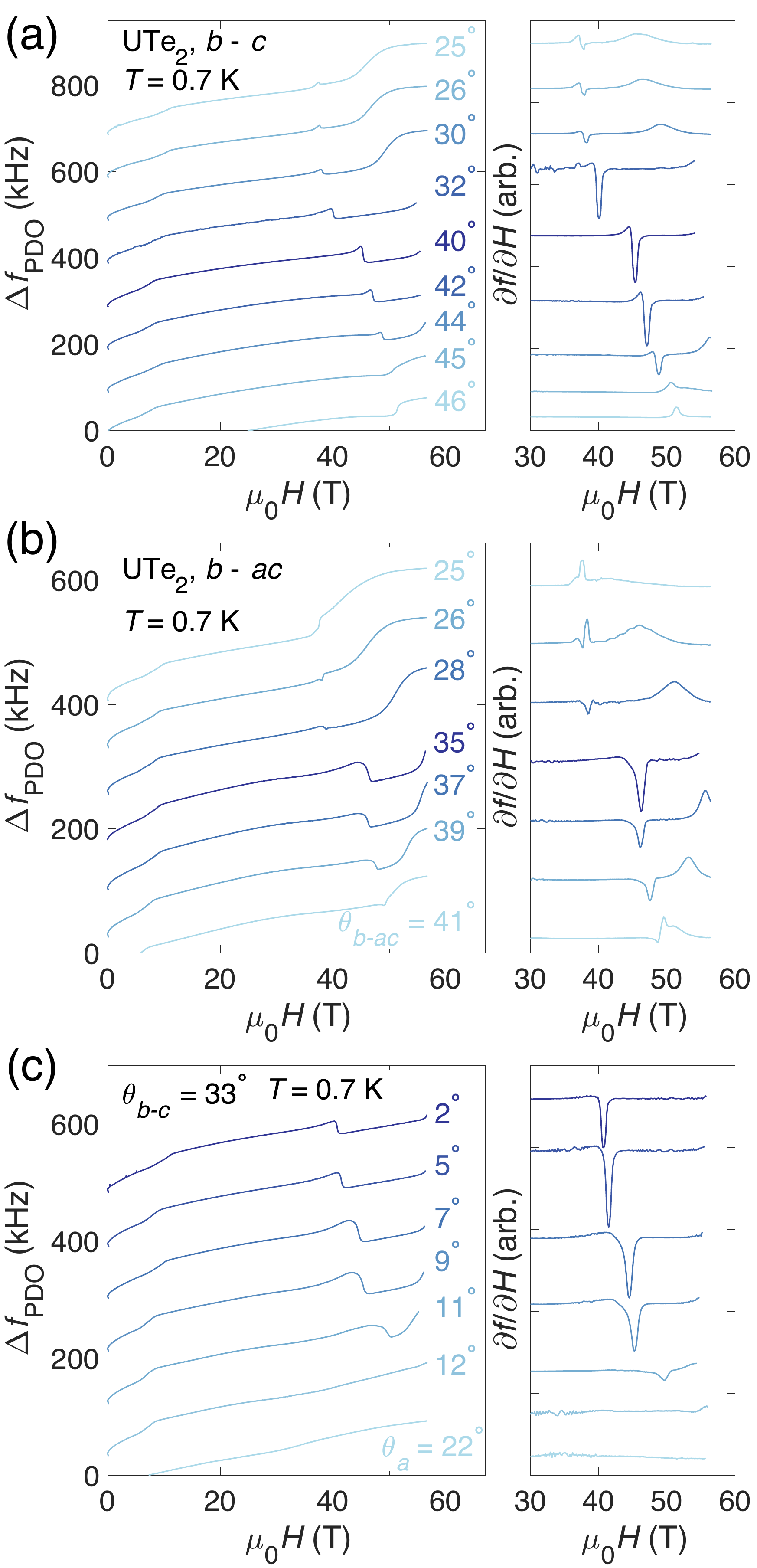} 
     \end{center}
     \vspace{-5pt}
     \caption{PDO measurements tracking the angular extent of the SC3 phase up to $\mu_0 H$~=~57~T at $T$~=~0.7~K in (a) the $b-c$ plane; (b) the $b-ac$ plane; (c) from $\theta_{b-c}$~=~33$\degree$ rotating towards $a$.}
    \label{pic:SC3-wuhan}
\end{figure}

\begin{figure*}[h!]
    \vspace{2cm}
     \begin{center}
     \includegraphics[width=.8\linewidth]{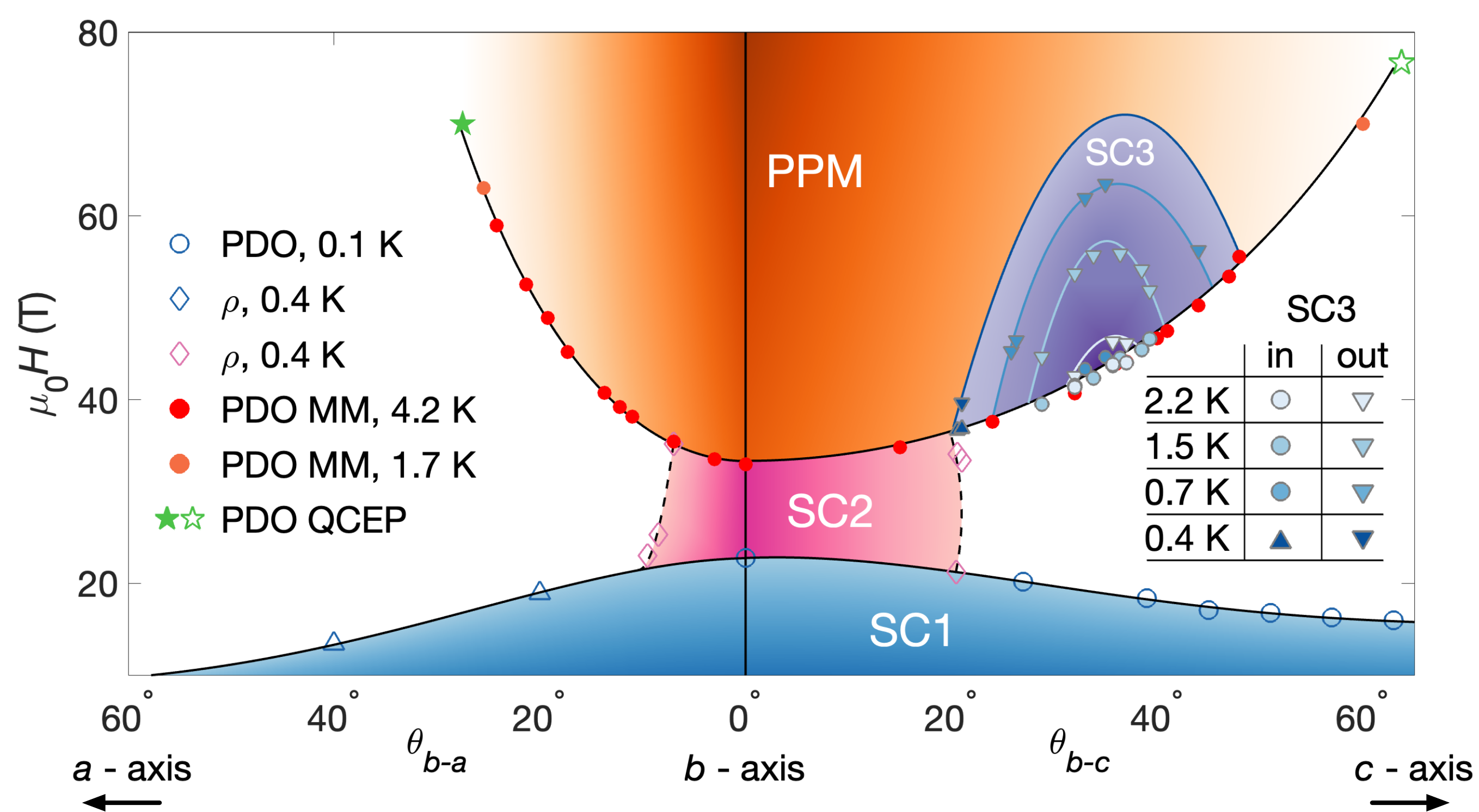} 
     \end{center}
     \vspace{-5pt}
     \caption{High field UTe$_2$ phase diagram for magnetic field orientations in the $bc$ and $ba$ rotational planes. The phase boundaries of the SC1, SC2, SC3 and PPM phases are determined by our prior measurements reported in refs.~\cite{tony2024enhanced,tony2025brief} and by those from this study.}
    \label{fig:phase-diag-2d}
\end{figure*}

\begin{figure}[h!]
    \vspace{0cm}
     \begin{center}
     \includegraphics[width=1\linewidth]{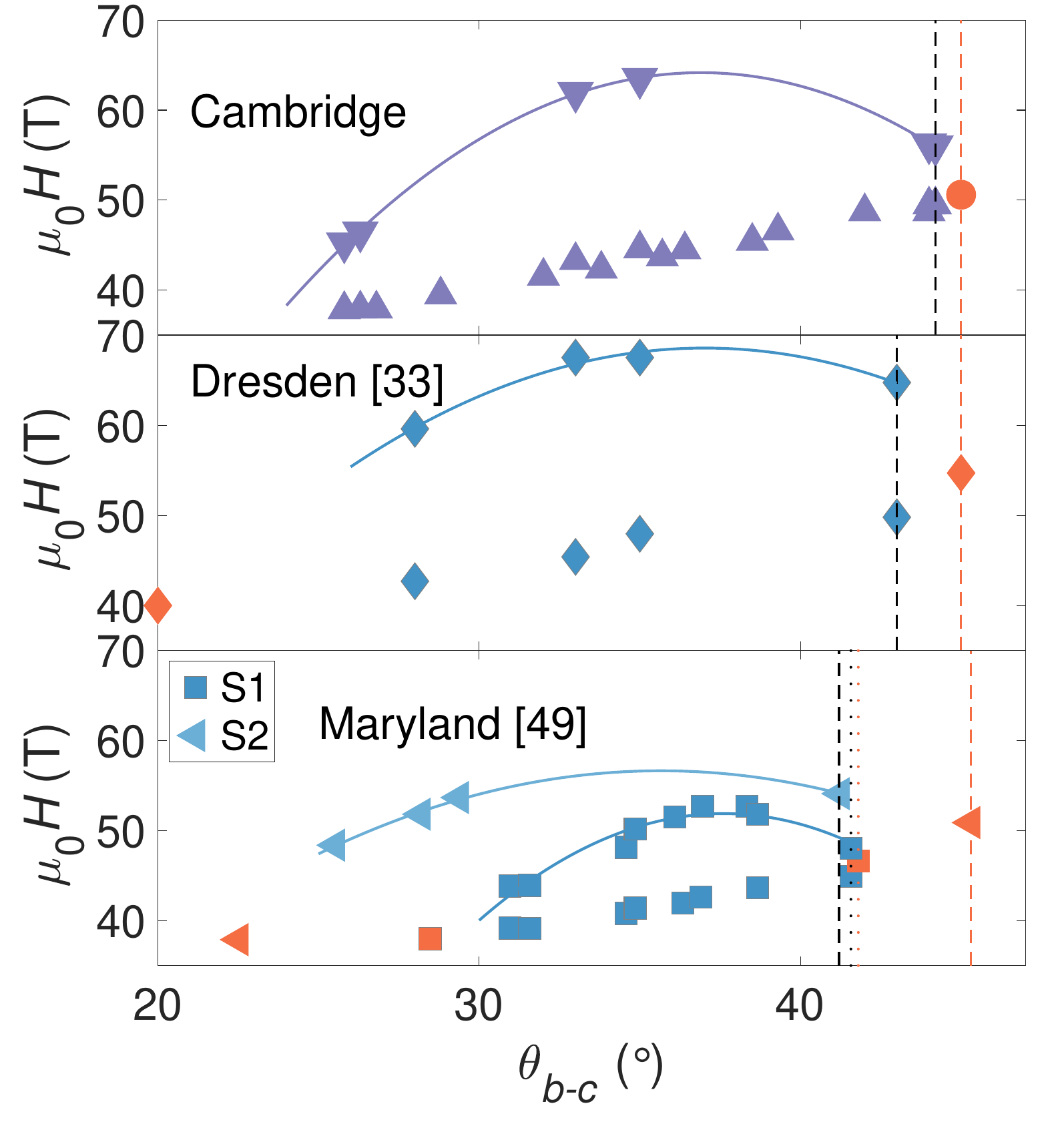} 
     \end{center}
     \vspace{-5pt}
     \caption{Comparison of the angular extent of SC3 in the $b-c$ rotation plane from measurements of electrical conductivity presented in this study along with those reported in refs.~\cite{helm2024,frank2024orphan}. All data are at $T \approx$~0.5-0.7~K. Orange points represent the transition into the PPM state, while purple and blue symbols indicate the start and end of the SC3 state. Black dashed lines mark the highest angle at which SC3 is still observed; orange dashed lines mark the next subsequent measured angle from each study, where signatures of superconductivity are no longer resolved (only the transition to the PPM phase). Note that two samples are plotted from ref.~\cite{frank2024orphan}. The SC3 state seems to terminate over a similar range of inclination of 41$\degree \lessapprox \theta_{b-c} \lessapprox$~45$\degree$ in each of these studies.}
    \label{fig:SC3comparison}
\end{figure}

\begin{figure}[h!]
    \vspace{1cm}
     \begin{center}
     \includegraphics[width=1\linewidth]{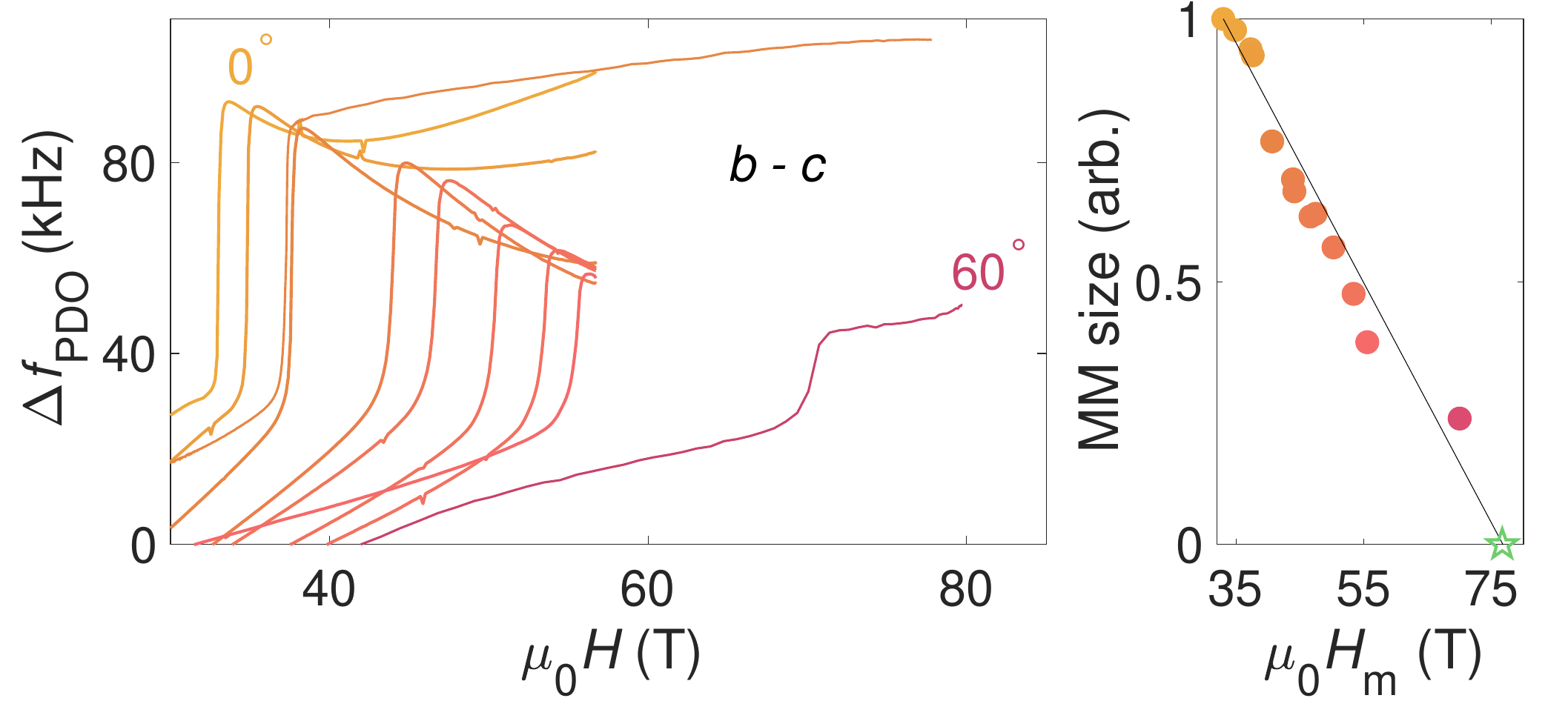} 
     \end{center}
     \vspace{-5pt}
     \caption{Extrapolation of the $b-c$ plane metamagnetic transition to its QCEP at high field. Note that Dresden data in Fig.~\ref{fig:MMdeath} has been rescaled by a factor of 0.57 to compare with Wuhan data, while here we plot data from Dresden measured on a different sample in a different coil that has been rescaled by a factor of 1.1, which gives the same size of MM jump at $\theta_{b-c}$~=~24$\degree$ (where $\mu_0 H_m =$~37.4~T). By taking the size of the metamagnetic transition as a function of $\theta_{b-c}$, we fit a linear trendline to extrapolate to the QCEP and plot open green stars at $|\mu_0 H_c|$~=~69~T ($\mu_0 H_m =$~76.7~T) in Fig.~\ref{fig:3D_PD}.}
    \label{fig:bc-extrpol}
\end{figure}

\clearpage

\begin{figure}[h!]
    \vspace{0cm}
     \begin{center}
     \includegraphics[width=1\linewidth]{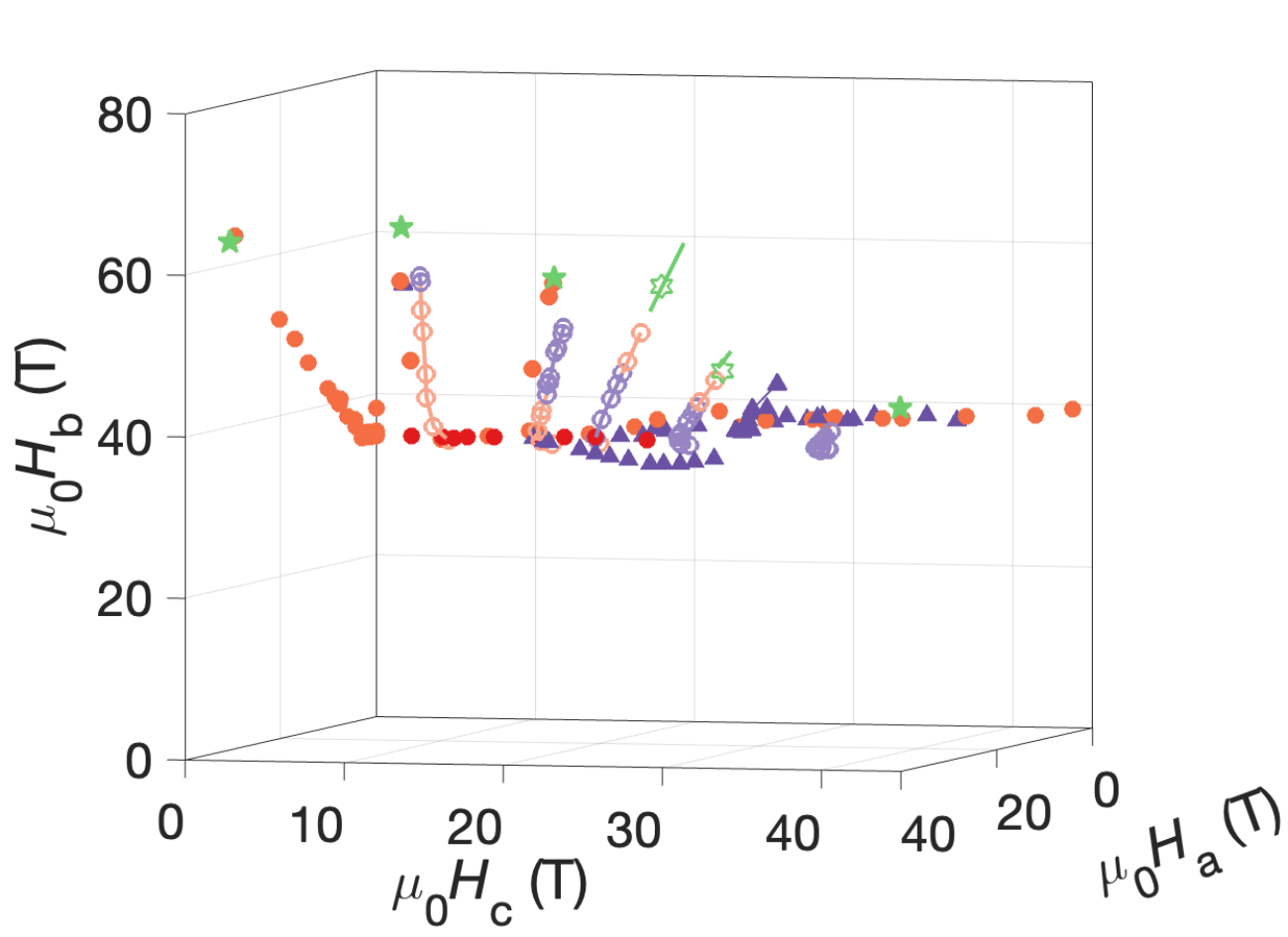} 
     \end{center}
     \vspace{-5pt}
     \caption{Comparison of high field PPM and SC3 data measured in this study with those reported in Lewin et al.~\cite{lewin2024halo}. Our data are plotted with solid symbols, while those of ref.~\cite{lewin2024halo} are given as open symbols. Solid green stars are QCEPs from our measurements, while open green hexagrams are interpreted from the data in ref.~\cite{lewin2024halo}. We plot an indicative error bar for the QCEPs interpreted from the data in Lewin et al.~\cite{lewin2024halo}, as those data were measured below $T_c^{SC3}$ therefore it is difficult to definitively determine the location of the QCEP as the domain of SC3 can stretch on either side of the QCL, as we show in Fig.~\ref{fig:SC3_Spill}. We therefore include these indicative error bars, constructed by spanning the angular region halfway between the last measurement point clearly showing first-order-like metamagnetism and the suppression towards 0~K. We find that the dataset reported in Lewin et al.~\cite{lewin2024halo} agrees very well with our construction of the UTe$_2$ high magnetic field QCL.}
    \label{pic:lewin-comp}
\end{figure}

\begin{figure}[h!]
    \vspace{1cm}
     \begin{center}
     \includegraphics[width=.75\linewidth]{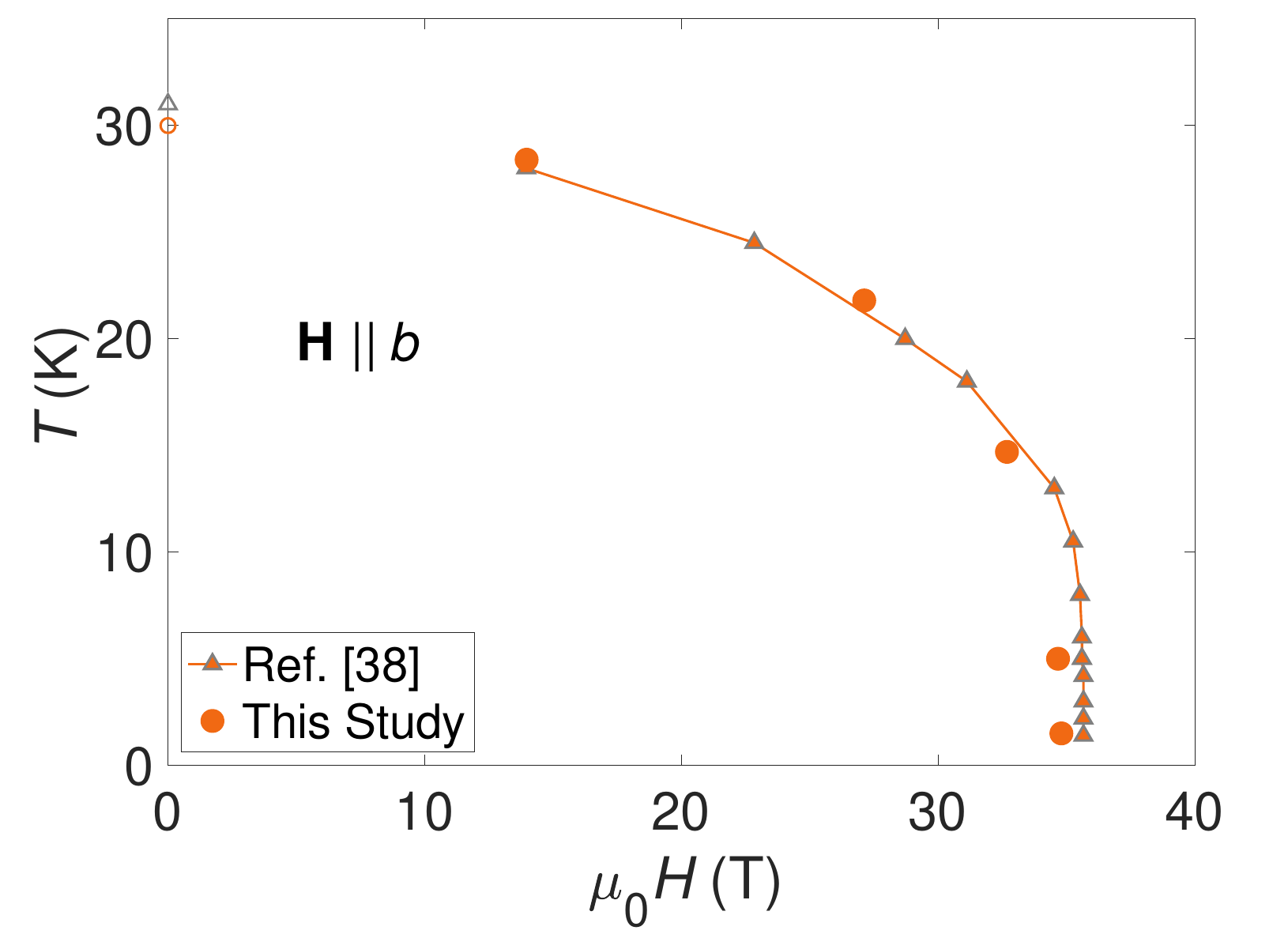} 
     \end{center}
     \vspace{-5pt}
     \caption{Temperature evolution of the metamagnetic transition for \textbf{H}~$\parallel b$. Circular points are from Fig.~\ref{pic:derivs} with triangular points from ref.~\cite{knafo2019magnetic}. The metamagnetic transition field $H_m$ decreases monotonically with elevated temperatures. Our observation of monotonically rising $\rho(H)$ (and $\nicefrac{\partial \rho}{\partial H}$) at $\theta_{b-ac} =$~35$\degree$ for $T =$~5~K in Fig.~\ref{pic:derivs} therefore shows that $\mu_0 H_m$ lies higher than 41.5~T. However, remarkably, at low temperatures we observe zero resistance at $\mu_0 H <$~41.5~T (Fig.~\ref{fig:SC3_Spill}), indicating the spilling over of SC3 beyond the PPM phase.}
    \label{pic:Tczoom}
\end{figure}

\clearpage


%

\end{document}